\pdfoutput=1

\documentclass[]{pasa}

\usepackage{graphicx,amssymb}
\usepackage{epsf,psfig}
\usepackage{natbib}
\usepackage[unicode]{hyperref}

\title[Introducing a new 3D dynamical model for barred galaxies]{Introducing a new 3D dynamical model for barred galaxies}
\author[Ch. Jung \& E. E. Zotos]{Christof Jung$^1$ \and Euaggelos E. Zotos$^{2,3}$\\
\affil{$^1$Instituto de Ciencias F\'{i}sicas,
Universidad Nacional Aut\'{o}noma de M\'{e}xico
Av. Universidad s/n, 62251 Cuernavaca, Mexico}
\affil{$^2$Department of Physics, School of Science,
Aristotle University of Thessaloniki,
GR-541 24, Thessaloniki, Greece}
\affil{$^3$Email: \href{mailto:evzotos@physics.auth.gr}{evzotos@physics.auth.gr}\\
\vspace*{1\baselineskip}
(RECEIVED May 23, 2015; ACCEPTED September 24, 2015)}}

\jid{PASA}
\volume{32, e042, 20 pages (2015).}
\doi{\href{http://dx.doi.org/10.1017/pasa.2015.43}{10.1017/pasa.2015.43}}
\jyear{\the\year}

\begin{document}

\begin{abstract}

The regular or chaotic dynamics of an analytical realistic three dimensional model composed of a spherically symmetric central nucleus, a bar and a flat disk is investigated. For describing the properties of the bar we introduce a new simple dynamical model and we explore the influence on the character of orbits of all the involved parameters of it, such as the mass and the scale length of the bar, the major semi-axis and the angular velocity of the bar as well as the energy. Regions of phase space with ordered and chaotic motion are identified in dependence on these parameters and for breaking the rotational symmetry. First we study in detail the dynamics in the invariant plane $z = p_z = 0$ using the Poincar\'e map as a basic tool and then we study the full 3 dimensional case using the SALI method as principal tool for distinguishing between order and chaos. We also present strong evidence obtained through the numerical simulations that our new bar model can realistically describe the formation and the evolution of the observed twin spiral structure in barred galaxies.
\end{abstract}


\maketitle

\section{INTRODUCTION}
\label{intro}

It is well known that the axial symmetry in galaxies is only a first approach. In essence, galaxies exhibit deviation from the axial symmetry which can be very small or more extended. In the latter category, we may include the case of barred galaxies. Observations indicate that a large percentage of disk galaxies, about 70\%, shows bar-like formations \citep[e.g.,][]{Ee00,SRSS03}. Bars are linear extended structures located in the central regions of galaxies. Usually bars are formed either from angular momentum redistribution within the disk or from disk instabilities \citep[e.g.,][]{A03,BSMH07,FCT08}.

\citet{K79} conducted the first systematic investigation regarding the sizes of the galactic bars. He found that the size of the bar is actually correlated with the luminosity of the galaxy. In the same vein, \citet{EE85} showed that bars in early-type disk galaxies tended to be larger than bars in later Hubble types \citep{RE97}. Recent observations using CCDs or near-infrared images supported the initial findings of Kormendy and Elmegreen \citep[e.g.,][]{CCD99,LSKP02,LS02,LSR02}. Moreover it is observed that barred galaxies may display different characteristics regarding the size of the bar. There are galaxies with a prominent barred structure \citep{EE85} and also galaxies with faint weak bars \citep{WK01}.

Galactic bars are very efficient at driving gas inwards and thus may help to frow central bulge components in galaxy disks \citep[e.g.,][]{DYB04,DMC06,G11}. Galaxies with earlier-type morphologies, which have more prominent bulges, tend to have more extended bars \citep[e.g.,][]{BJM08,HMN11,MNH11,WJK09}. Furthermore, in some galaxies, the central bulges and the bars seem to contain similar stellar populations \citep{SBO11}. However, there are some barred galaxies that lack completely bulges and many bulge-dominated galaxies which lack bars \citep[e.g.,][]{LSBK07,PSB11}. It should also be noted that in some cases, the formation of the central bulges is the result caused by disk dynamical instabilities \citep{KK04}.

An important phenomenon in barred galaxies is associated with rings which are active sites of new star formation \citep[e.g.,][]{KBHSd95,MKVR08,Se10,HMLHOW11}. It is believed that the formation of nuclear rings is due to the effect of a non axially symmetric potential of the bar in a large supply of interstellar gas. A key role in this mechanism is played by the torque of the bar, which causes the gas to form the nuclear rings \citep{KSSYT12}. Observations show that the rate of star formation in the nuclear rings not only is different in several types of barred galaxies but also varies significantly with time \citep[e.g.,][]{BTBC00,BHJKS02,CKB10}.

The formation and evolution of dust lanes and nuclear rings have been extensively studied using numerical simulations \citep[e.g.,][]{PST95,EG97,MTSS02,RT03,TAJ09}. The formation of nuclear rings from the resonant interaction of the gas with the potential of the bar appears not to be consistent with recent studies, suggesting the action of a different mechanism \citep[e.g.,][]{KSK12}. According to this mechanism, there is a centrifugal barrier which cannot be overcome by the inflowing gas. This barrier is responsible for the formation of the nuclear rings. Finally recent research reveals that more massive
bars cause smaller nuclear rings, for supporting observational data see \citep{CKB10}.

Over the last decades, a huge amount of research work has been devoted to understanding the orbital structure in barred galaxy models \citep[e.g.,][]{ABMP83,P84,CDFP90,A92,P96,KC96,OP98,PMM04}. The reader can find more information about the dynamics of barred galaxies in the reviews by \citet{A84,CG89,SW93}. We would like to point out, that all the above-mentioned references on the dynamics of barred galaxies are exemplary rather than exhaustive. However, we should like to discuss briefly some of the recent papers on this subject. \citet{SPA02a} conducted an extensive investigation regarding the stability and morphology of both 2D and 3D periodic orbits in a fiducial model representative of a barred galaxy. The work was continued in the same vein in \citet{SPA02b}, where the influence of the system's parameters on the 3D periodic orbits was revealed. Moreover, \citet{KP05} presented evidence that in two-dimensional models with sufficiently large bar axial ratios, stable orbits having propeller shapes play a dominant role to the bar structure. \citet{MA11} estimated the fraction of chaotic and regular orbits in both two and three-dimensional potentials by computing several sets of initial conditions and studying how these fractions evolve when the energy and also basic parameters of the model, such as the mass, size and pattern speed of the bar vary. Computing the statistical distributions of sums of position coordinates \citet{BMA12} quantified weak and strong chaotic orbits in 2D and 3D barred galaxy models. A time-dependent barred galaxy model was utilized in \citet{MBS13} in order to explore the interplay between chaotic and regular behaviour of star orbits when the parameters of the model evolve.

So far many publications use the Ferrer's triaxial model \citep{F77} to describe the bar. The corresponding potential however is too complicated and it is not known in a closed form. On this basis, we decided to construct a new simpler but non the less realistic potential with a clear advantage on the performance speed of the numerical calculations in comparison with the Ferrer's potential.

The paper is organized as follows: in Section \ref{galmod}, we present the structure and the properties of the new barred galaxy model. In section \ref{numres2} we construct Poincar\'e maps in order to investigate how all the parameters entering the bar model influence the orbital properties in the invariant surface $z = p_z = 0$. In section \ref{numres3} we use the SALI method \citep{S01} to reveal the influence of the parameters on the full 3D properties. In the following section we present numerical evidence that our dynamical model can realistically simulate the creation as well as the evolution of twin spiral arms of a real barred galaxy. The paper ends with Section \ref{disc}, where the discussion and the main conclusions of our work are presented.

\section{THE NEW DYNAMICAL MODEL}
\label{galmod}

Our total analytical gravitational three-dimensional (3D) potential $\Phi(x,y,z)$ consists of three main components: a central spherical component $\Phi_{\rm n}$, a flat disk $\Phi_{\rm d}$ and a bar potential $\Phi_{\rm b}$. We decided to include only these three components, so as to be able to directly relate our results with those of the corresponding three component model presented in \citet{P84}.

For the description of the spherically symmetric nucleus we use a Plummer potential \citep{BT08}
\begin{equation}
\Phi_{\rm n}(x,y,z) = - \frac{G M_{\rm n}}{\sqrt{x^2 + y^2 + z^ 2 + c_{\rm n}^2}},
\label{Vn}
\end{equation}
where $G$ is the gravitational constant, while $M_{\rm n}$ and $c_{\rm n}$ are the mass and the scale length of the nucleus, respectively. At this point we must point out that potential (\ref{Vn}) is not intended to represent a black hole nor any other compact object, but a dense and massive bulge (nucleus). Therefore, we do not include any relativistic effects.

The flat disk is modelled by a Miyamoto-Nagai potential \citep{MN75}
\begin{equation}
\Phi_{\rm d}(x,y,z) = - \frac{G M_{\rm d}}{\sqrt{x^2 + y^2 + \left(k + \sqrt{h^2 + z^ 2}\right)^2}},
\label{Vd}
\end{equation}
where $M_{\rm d}$ is the mass of the disk, while $k$ and $h$ are the horizontal and vertical scale lengths of the disk, respectively.

In order to model the properties of the galactic bar we decided to develop a new dynamical model. The basic idea is to construct a bar from a mass distribution of length $2a$ which lies along the $x$-axis and has a continuum of centers going from $x = -a$ to $x = +a$. We call this interval lying in the three dimensional position space $I_B$. The total mass of the bar is $M_{\rm b}$. We imagine that in a first step we have a mass distribution along $I_B$ with a constant linear mass density $m = M_{\rm b}/2a$. To avoid all problems with singularities of point densities and in order to include the transverse width $c_{\rm b}$ of the bar we imagine that in a second step the linear density $m$ belonging to a single point of $I_B$ is smeared out over the effective width $c_{\rm b}$. Thereby each point $s$ from the interval $I_B$ becomes the center of the regularized monopole potential of the form
\begin{equation}
v(x-s,y,z) = - \frac{G m}{ \sqrt{(x-s)^2 + d^2}},
\label{monop}
\end{equation}
where
\begin{equation}
d^2 = y^2 + z^2 + c_b^2
\end{equation}
and the total potential $\Phi_{\rm b}(x,y,z)$ becomes the integral of these contributions over the interval $I_B$, i.e.
\[
\Phi_{\rm b}(x,y,z) = \int^{+a}_{-a} ds \; \;   v(x-s,y,z)
\]
\[
= \frac{G M_{\rm b}}{2a}\left[ arcsinh \left( \frac{x-a}{d} \right) - arcsinh \left( \frac{x+a}{d} \right) \right]
\]
\begin{equation}
= \frac{G M_{\rm b}}{2a} \ln \left( \frac{x-a+\sqrt{(x-a)^2 + d^2}} {x+a+\sqrt{(x+a)^2 + d^2}} \right).
\label{Vb}
\end{equation}
As always we recover the corresponding mass density $\rho(x,y,z)$ from the Poisson's equation
\[
\rho(x,y,z) = \frac{1}{4 \pi G} \nabla^2 \Phi_{\rm b}(x,y,z)
\]
\[
= \frac{M_{\rm b} c_{\rm b}^2}{8 \pi a} \int^{+a}_{-a} ds ((x-s)^2 + d^2)^{-5/2}
\]
\begin{equation}
= \frac{M_{\rm b} c_{\rm b}^2}{8 \pi a} (g(x+a,y,z) - g(x-a,y,z)),
\end{equation}
where
\begin{equation}
g(x,y,z) = x(2x^2 + 3d^2)d^{-2}(x^2 + d^2)^{-3/2}.
\end{equation}
Note that the integral of $\rho$ over all space gives correctly $M_{\rm b}$. The rotationally symmetric limit  $a \to 0$ gives correctly
\begin{equation}
\Phi_{\rm b}(x,y,z) = - \frac{G M_{\rm b}}{ \sqrt{x^2 + d^2}},
\end{equation}
\begin{equation}
\rho(x,y,z) = \frac{3 M_{\rm b} c_{\rm b}^2}{4 \pi} (x^2 + d^2)^{-5/2}
\end{equation}

The bar rotates clockwise around its $z$-axis at a constant angular velocity $\Omega_{\rm b}$. The major axis of the bar points into the $x$ direction while its intermediate axis points into the $y$ direction. Therefore the effective potential is
\begin{equation}
\Phi_{\rm eff}(x,y,z) = \Phi(x,y,z) - \frac{1}{2}\Omega_{\rm b}^2 \left(x^2 + y^2 \right).
\label{Veff}
\end{equation}

We use a system of galactic units, where the unit of length is 1 kpc, the unit of mass is $2.325 \times 10^7 {\rm M}_\odot$ (solar masses) and the unit of time is $0.9778 \times 10^8$ yr (about 100 Myr). The velocity unit is 10 km/s, the unit of angular momentum (per unit mass) is 10 km kpc s$^{-1}$, while $G$ is equal to unity. The energy unit (per unit mass) is 100 km$^2$s$^{-2}$, while the angle unit is 1 radian. In these units, the values of the involved parameters are: $M_{\rm n} = 400$ (corresponding to 9.3 $\times 10^{9}$ ${\rm M}_\odot$), $c_{\rm n} = 0.25$, $M_{\rm d} = 7000$ (corresponding to 1.6275 $\times 10^{11}$ ${\rm M}_\odot$), $k = 3$, $h = 0.175$, $M_{\rm b} = 3500$ (corresponding to 8.13 $\times 10^{10}$ ${\rm M}_\odot$), $a = 10$, $c_{\rm b} = 1$ and $\Omega_{\rm b} = 1.25$. This set of the values of the parameters defines the Standard Model (SM).

\begin{figure}
\begin{center}
\includegraphics[width=\hsize]{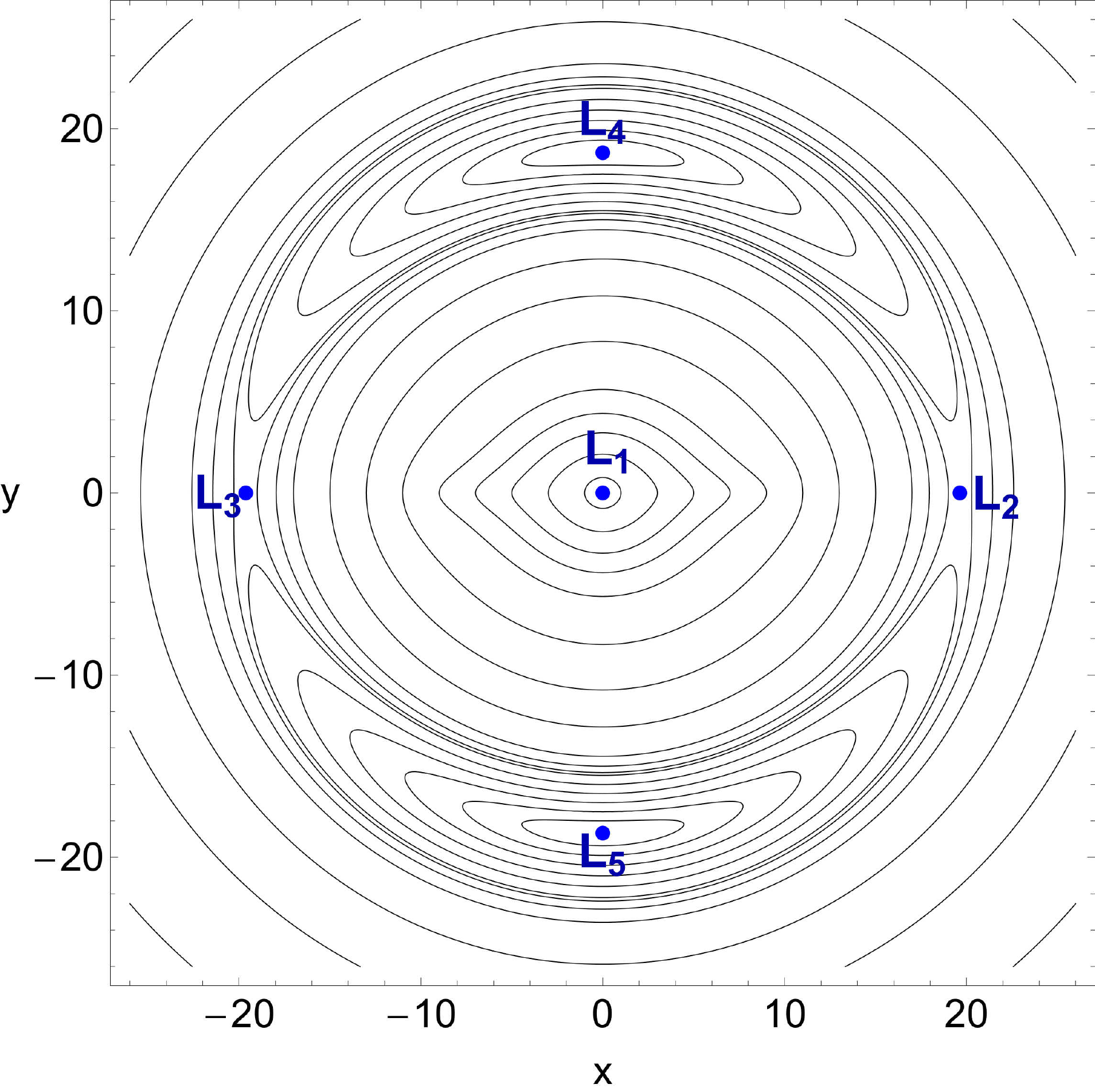}
\end{center}
\caption{The isoline contours of the effective potential in the $(x,y)$-plane for $z = 0$ for the standard model. Included are the five Lagrangian points.}
\label{isoc}
\end{figure}

\begin{figure}
\begin{center}
\includegraphics[width=\hsize]{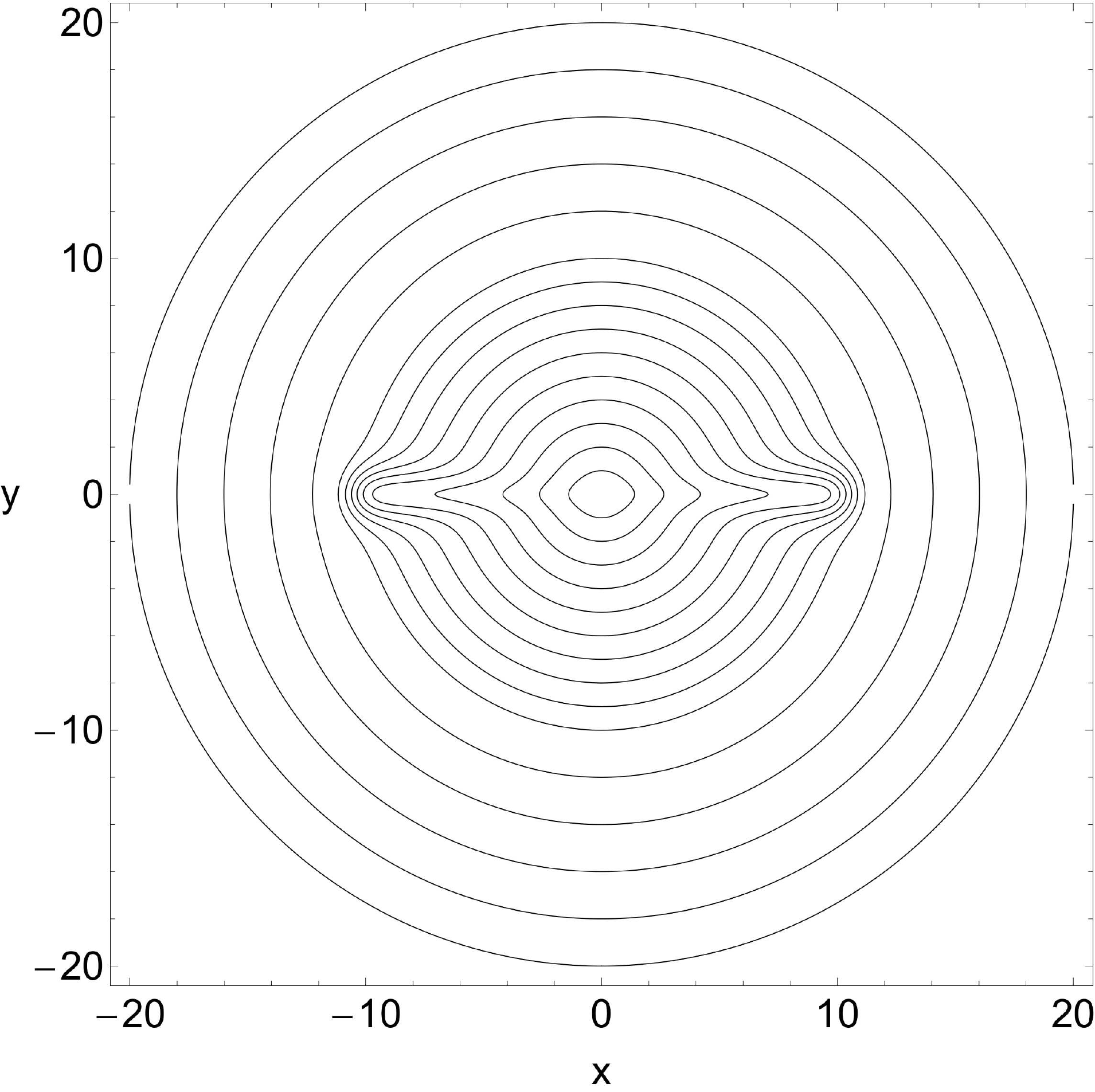}
\end{center}
\caption{The isoline contours of the mass density on the $(x,y)$-plane for $z = 0$ of the total gravitational potential $\Phi(x,y,z)$ for the standard model.}
\label{isod}
\end{figure}

The isoline contours of constant effective potential on the $(x,y)$-plane for $z = 0$ as well as the position of the five Lagrangian points $L_i, \ i = {1,5}$ are shown in Fig. \ref{isoc}. Three of them, $L_1$, $L_2$, and $L_3$, known as the collinear points, are located on the $x$-axis. The central stationary point $L_1$ at $(x,y) = (0,0)$ is a local minimum of $\Phi_{\rm eff}$. At the other four Lagrangian points it is possible for the test particle to move in a circular orbit, while appearing to be stationary in the rotating frame. For this circular orbit, the centrifugal and the gravitational forces precisely balance. The stationary points $L_2$ and $L_3$ are saddle points, while points $L_4$ and $L_5$ on the other hand, are local maxima of the effective potential, enclosed by the banana-shaped isolines. The annulus bounded by the circles through $L_2$, $L_3$ and $L_4$, $L_5$ is known as the ``region of coroation". The numerical value of the effective potential at the two Lagrangian points $L_2$ and $L_3$ results in a critical Jacobi constant $C_L = -869.70514693$. Similarly, in Fig. \ref{isod} we observe the isoline contours of the mass density on the $(x,y)$-plane for $z = 0$ of the total gravitational potential. The barred structure is clearly visible in the region $-10 \leq x \leq +10$ kpc. Here we should emphasize that the shape of the equidensity curves is very similar to that of the Ferrer's ellipsoid (see e.g., Fig. 1 in \citet{P84})

The motion of a test particle with a unit mass $(m = 1)$ in our rotating barred galaxy model is governed by the Hamiltonian
\begin{equation}
H = \frac{1}{2} \left(p_x^2 + p_y^2 + p_z^2 \right) + \Phi(x,y,z) - \Omega_{\rm b} \left( x p_y - y p_x\right) = E,
\label{ham}
\end{equation}
where $p_x$, $p_y$ and $p_z$ are the momenta per unit mass, conjugate to $x$, $y$ and $z$ respectively, while $E$ is the numerical value of the Jacobi integral, which is conserved.

\begin{eqnarray}
\dot{x} &=& p_x + \Omega_{\rm b} y, \nonumber \\
\dot{y} &=& p_y - \Omega_{\rm b} x, \nonumber \\
\dot{z} &=& p_z, \nonumber \\
\dot{p_x} &=& - \frac{\partial \Phi}{\partial x} + \Omega_{\rm b} p_y, \nonumber \\
\dot{p_y} &=& - \frac{\partial \Phi}{\partial y} - \Omega_{\rm b} p_x, \nonumber \\
\dot{p_z} &=& - \frac{\partial \Phi}{\partial z},
\label{eqmot}
\end{eqnarray}
where the dot indicates derivative with respect to the time.

In the same vein, the variational equations which govern the evolution of a deviation vector ${\bf{w}} = (\delta x, \delta y, \delta z, \delta p_x, \delta p_y, \delta p_z)$ are
\begin{eqnarray}
\dot{(\delta x)} &=& \delta p_x + \Omega_{\rm b} \delta y, \nonumber \\
\dot{(\delta y)} &=& \delta p_y - \Omega_{\rm b} \delta x, \nonumber \\
\dot{(\delta z)} &=& \delta p_z, \nonumber \\
(\dot{\delta p_x}) &=&
- \frac{\partial^2 \Phi}{\partial x^2} \ \delta x
- \frac{\partial^2 \Phi}{\partial x \partial y} \delta y
- \frac{\partial^2 \Phi}{\partial x \partial z} \delta z + \Omega_{\rm b} \delta p_y, \nonumber \\
(\dot{\delta p_y}) &=&
- \frac{\partial^2 \Phi}{\partial y \partial x} \delta x
- \frac{\partial^2 \Phi}{\partial y^2} \delta y
- \frac{\partial^2 \Phi}{\partial y \partial z} \delta z - \Omega_{\rm b} \delta p_x, \nonumber \\
(\dot{\delta p_z}) &=&
- \frac{\partial^2 \Phi}{\partial z \partial x} \delta x
- \frac{\partial^2 \Phi}{\partial z \partial y} \delta y
- \frac{\partial^2 \Phi}{\partial z^2} \delta z
\label{vareq}
\end{eqnarray}

\section{NUMERICAL RESULTS FOR THE 2-DOF SYSTEM}
\label{numres2}

The subspace S$_z$ of the phase space defined by $z = 0$ and $p_z = 0$ is invariant, i.e. initial conditions in $S_z$ lead to orbits lying entirely in $S_z$. The restriction of the dynamics to $S_z$ will be the restricted system with two degrees of freedom. We frequently use the standard polar coordinates $R$ and $\phi$ in the $(x,y)$-plane and their conjugate momenta $p_R$ and $L$.

The restricted system has several integrable limit cases. The most important ones are the limits $a = 0$ and $M_{\rm b} = 0$. Both of these cases are rotationally symmetric and therefore the intersection condition of the Poincar\'{e} map and the variables used in the domain of the map should be chosen with these limit cases in mind to describe well these limit cases, where the angular degree of freedom is decoupled from the rest of the system (this also holds for the full 3-dof system). Ideally the symmetric limit cases should be described by pure twist maps and their corresponding twist curves. We study the restriction of the Poincar\'{e} map to the invariant subset $z = p_z = 0$ which will be called $P_r$ in the following and it is a pure twist map in the rotationally symmetric limit cases. Therefore the variables used in the full dimensional Poincar\'{e} map should contain the action-angle variables of this degree of freedom, i.e. $L$ and $\phi$ and for the restricted map the domain is simply the cylindrical domain of these two coordinates. Then naturally the intersection condition must be a condition in the variables $R$ and $p_R$. At the moment we are only interested in bound motion and then the most simple and natural intersection condition is the condition that $R$ runs through a relative maximum and $p_R$ changes from positive to negative values. Almost all bound orbits run an infinite number of times through this intersection condition in the limits $t \to \pm \infty$. The only exceptions might be orbits with $R = constant$, e.g. the orbit in the 3-dof system which oscillates up and down the $z$-axis. And in the symmetric limit case there is the circular orbit around the origin with constant $R$. Fortunately, such orbits do not appear in the restricted system as soon as the rotational symmetry is broken. Therefore this intersection condition is the best one possible for our purposes.

\begin{figure*}
\centering
\resizebox{\hsize}{!}{\includegraphics{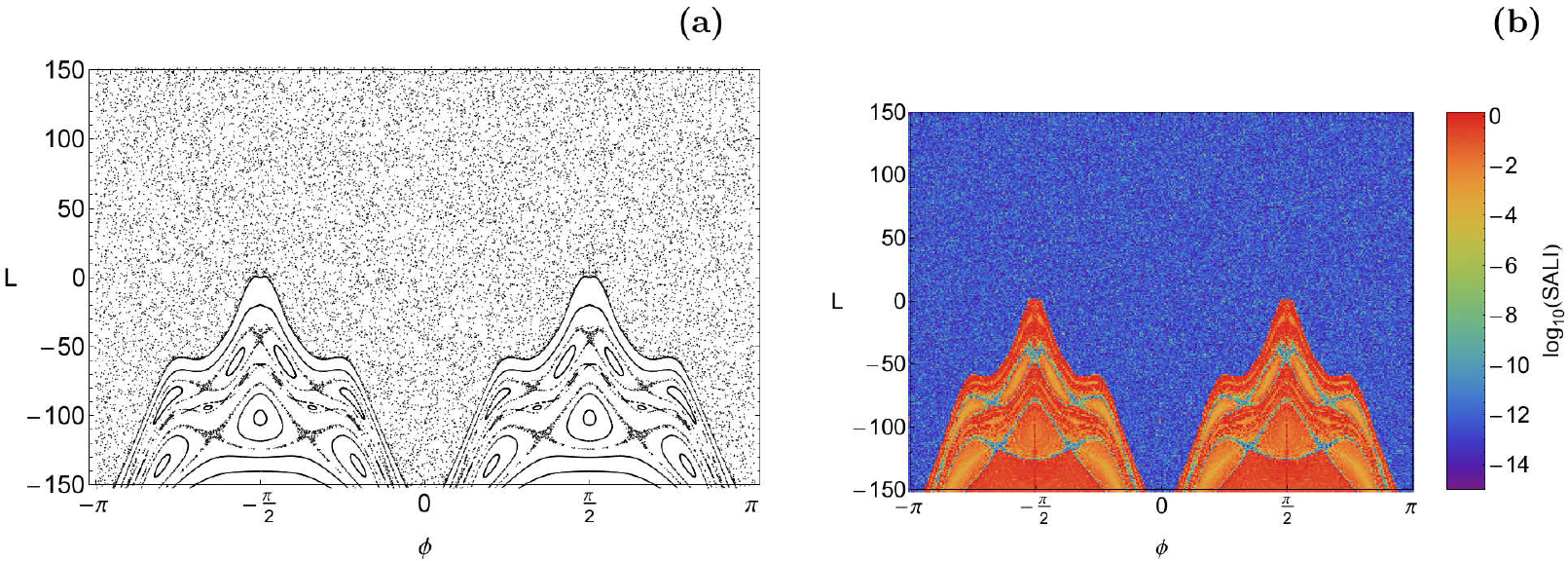}}
\caption{(a-left): The restricted Poincar\'{e} map $P_r$ for the standard model and (b-right): Regions of different values of the SALI in a dense grid of initial conditions on the $(\phi,L)$-plane. Light reddish colors correspond to ordered motion, dark blue/purple colors indicate chaotic motion, while all intermediate colors suggest sticky orbits. Note the excellent agreement between the two methods. However we should point out that the SALI method can easily pick out small stability regions embedded in the chaotic sea which cannot be easily detected by the classical PSS method.}
\label{sm}
\end{figure*}

To understand the dynamics of the restricted system we first pick a case which we use as standard case and then study the development of $P_r$ under the change of the various parameters in the potential of the bar. i.e. the parameters $M_{\rm b}$, $a$, $c_{\rm b}$, $\Omega_{\rm b}$ and finally we also are interested in the dependence on the value $E$ of the Hamiltonian in the rotating frame. As standard case we use $E = -900$, $M_{\rm b} = 3500$, $a = 10$, $c_{\rm b} = 1$, and $\Omega_{\rm b} = 1.25$. The corresponding plot of the restricted Poincar\'{e} map $P_r$ is presented in Fig. \ref{sm}a. In this figure and in the following ones we restrict the plot to the interval $L \in (-150,150)$ since this is the most relevant interval for the real galaxy. However, when useful for the understanding of the dynamics we will mention structures outside of this interval. Fig. \ref{sm}b shows the corresponding final SALI values obtained from the selected grid of initial conditions, in which each point is colored according to its SALI value at the end of the numerical integration (Please see Section \ref{numres3} for a description of our use of the SALI method). In this SALI plot light-reddish colors indicate ordered orbits, dark blue/purple colors correspond to chaotic orbits, while all the intermediate colors represent initial conditions of orbits whose chaotic nature is revealed only after long times (the so-called ``sticky orbits"\footnote{A sticky orbit is a chaotic orbit which behaves as a regular one for a long time period before revealing its true chaotic nature.}). A comparison of parts (a) and (b) of Fig. \ref{sm} demonstrates in which form SALI plots identify regions in the plane of initial conditions covered by regular motion. More details on the SALI method will be given in the beginning of the following section.

In the standard case the major part of the domain of the map is covered by a large scale chaotic sea, only the part for large negative values of $L$ shows mainly regular motion. The center of the regular motion is an orbit of period two at $\phi = \pm \pi/2$, $L \approx -170$. The corresponding orbit in position space encircles the origin in an almost circular orbit in negative (clockwise) orientation where the maximal values of $R$ lie along the $y$-axis (i.e. $\phi = \pm \pi/2$) creating the two corresponding points of period 2 in the map. As seen in the figure, from the secondary structures
around this basic stable periodic orbit the one of coupling ratio 1:3 is the most prominent one.

As usual for bound systems with a smooth Hamiltonian the large chaotic sea contains an infinity of small stable islands. However, the size of such islands and even the sum of all their sizes are so small that they do not have any significant influence on the global dynamical behaviour of the system. For all practical purposes we can treat the large chaotic sea as if it would be completely chaotic. This consideration also holds for all large scale chaotic seas which we encounter in the following Poincar\'{e} plots.

\subsection{Dependence on the mass of the bar}
\label{mass}

\begin{figure}
\begin{center}
\includegraphics[width=\hsize]{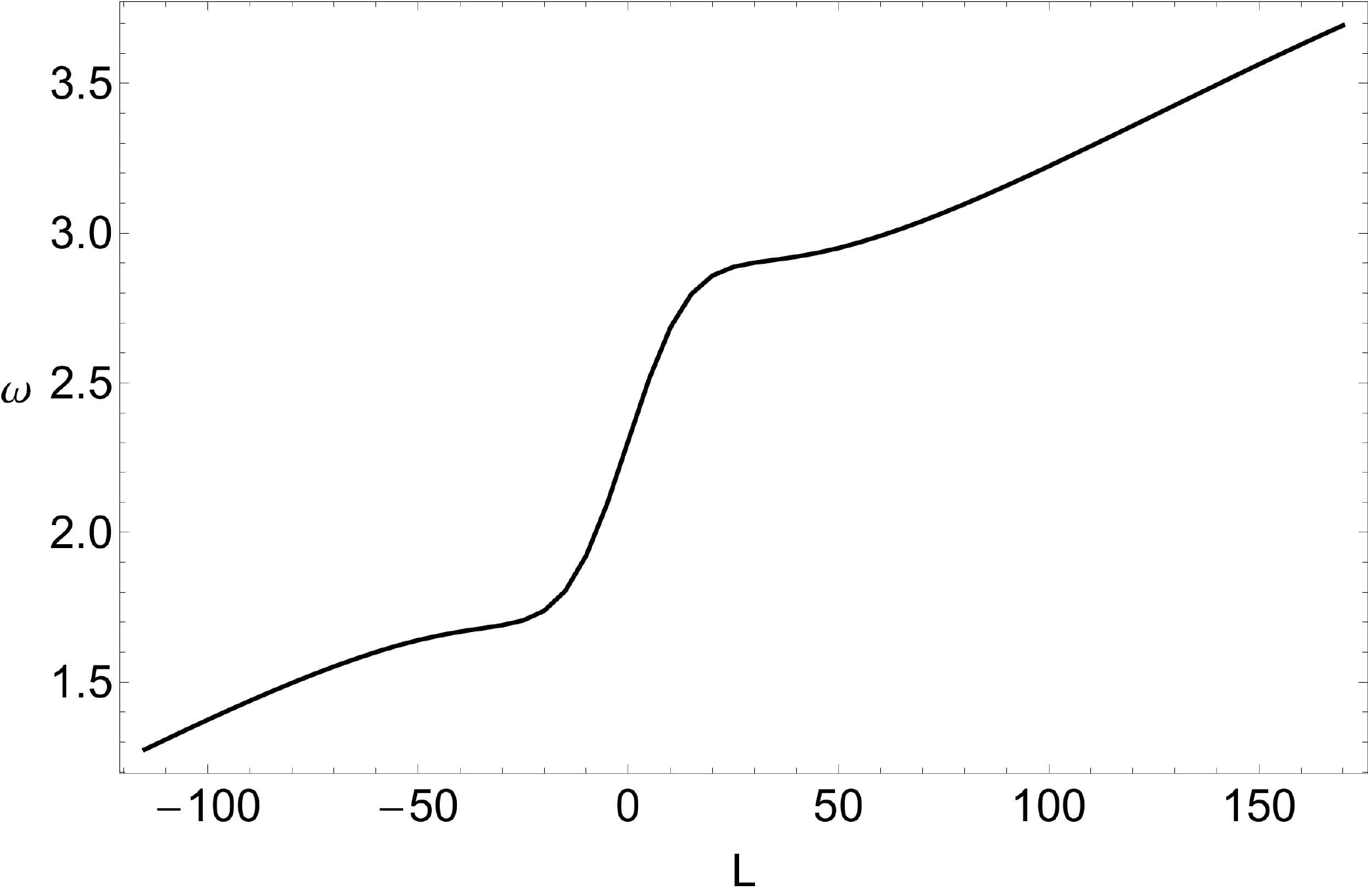}
\end{center}
\caption{The rotation angle $\omega$ as a function of the angular momentum $L$ for $M_{\rm b} = 0$.}
\label{mb0}
\end{figure}

\begin{figure*}
\centering
\resizebox{\hsize}{!}{\includegraphics{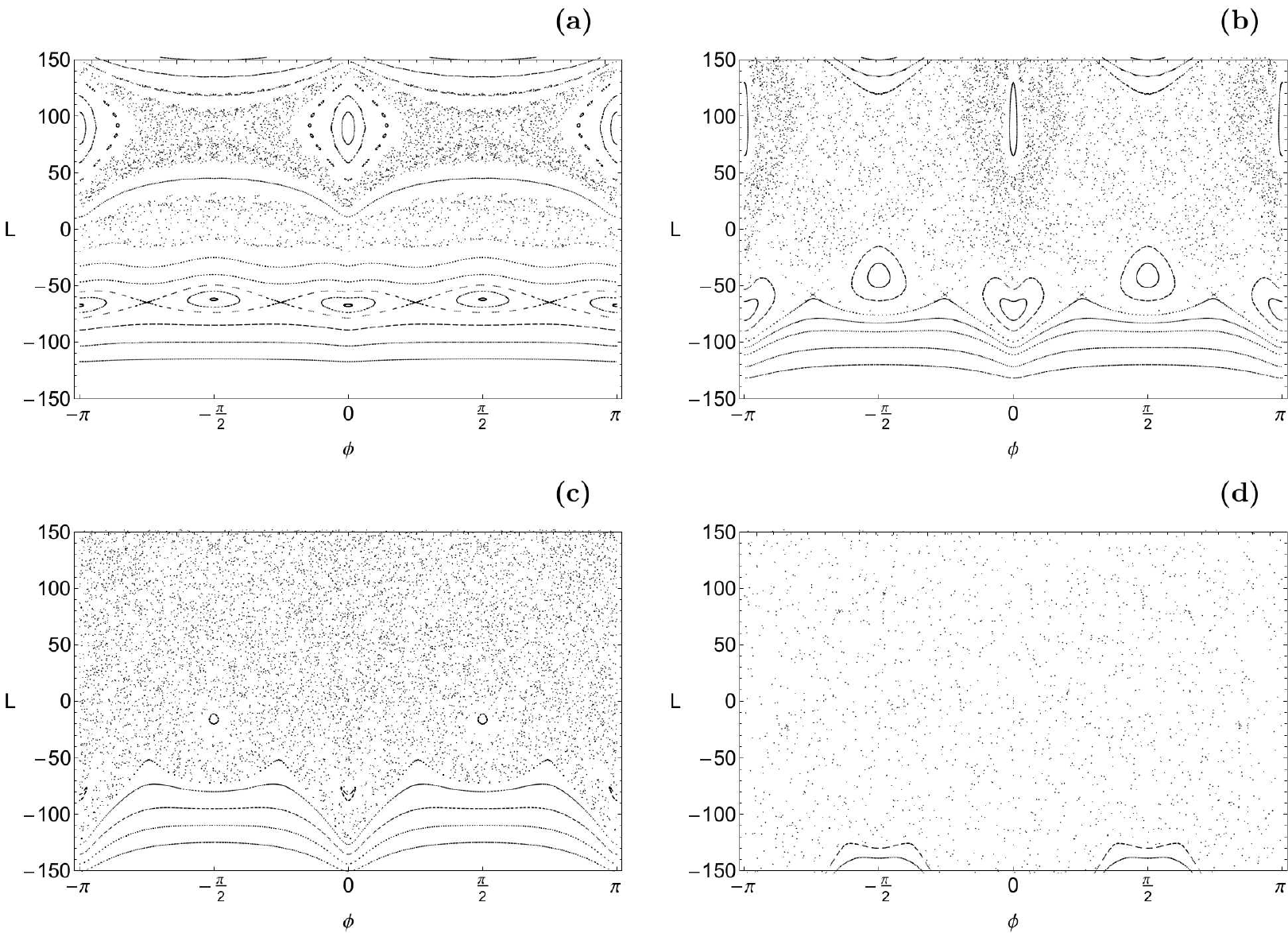}}
\caption{Examples of the perturbed map $P_r$ for various values of the mass of the bar $M_{\rm b}$ when $E = -900$. (a-upper left): $M_{\rm b} = 100$; (b-upper right): $M_{\rm b} = 500$; (c-lower left): $M_{\rm b} = 1000$; (d-lower right): $M_{\rm b} = 5000$.}
\label{mb}
\end{figure*}

In the case $M_{\rm b} = 0$ the bar has zero mass and therefore does not have any effect on the dynamics, i.e. the dynamics does not have any bar at all. Accordingly the dynamics is rotationally symmetric, $L$ is conserved and $P_r$ becomes a pure twist map. It is characterized by the rotation angle $\omega$ as function of $L$. In the position space of the rotating frame the twist angle omega gives the angular advance of the orbit between two consecutive relative maxima of the radial coordinate. This twist curve is plotted in Fig. \ref{mb0}. Note that for the integrable case $M_{\rm b} = 0$ the values of $a$ and $c_{\rm b}$ are irrelevant whereas these values matter for the details of the perturbation scenario away from the symmetric case. Under perturbations, i.e. here for $M_{\rm b}$ becoming different from zero, we expect $k:n$ resonances to become important whenever the twist angle has the value $2 \pi k/n$ for some small integer n. From the Fig. \ref{mb0} we expect a 1:4 resonance for $L \approx -70$, a 1:3 resonance for $L \approx 0$ and a 1:2 resonance for $L \approx 90$. In Fig. \ref{mb}(a-d) we show the perturbed map for four different nonzero values of $M_{\rm b}$. For the small value $M_{\rm b} = 100$ in part (a) we clearly see the large secondary 1:2 structure at a $L$ interval around the expected value of approximately 90. We observe secondary KAM islands centered around the period 2 elliptic points at angle values $\phi = 0$ and $\phi = \pi$ and a chaos strip organized by the hyperbolic period 2 points sitting at angle values $\phi = \pm \pi /2$. We also see clearly the 1:4 secondary structure around the $L$ value -70. Here the fine chaos strips organized by the hyperbolic points still come very close to a separatrix curve. The expected 1:3 resonance structure has already decayed into a chaos strip even for the small value $M_{\rm b} = 100$. As seen from Fig. \ref{mb0} the twist curve has a very large slope when it passes the value $2 \pi/3$ and therefore the corresponding resonance structure is extremely unstable and turns into a large scale chaos strip at very small perturbations. For values of $M_{\rm b} < 50$ we see two different chains of very small secondary islands of period 3. Because of the discrete symmetry $\phi \to \phi + \pi$ in our system the periodic points of odd period come in two different copies in general.

For the value $M_{\rm b} = 500$ in part (b) of the figure we still see the secondary KAM islands of the 1:2 resonance and also of the 1:4 resonance. The corresponding chaos strips have already merged to one global chaotic sea. For this value of $M_{\rm b}$ we still find primary KAM curves for large positive and large negative values of $L$. As seen in part (c) of the figure the secondary islands of the 1:2 resonance have disappeared for $M_{\rm b} = 1000$ whereas the secondary 1:4 islands still exist on a very small scale. For this perturbation value also the primary KAM curves for large values of $L$ have decayed. For the case of the very large value $M_{\rm b} = 5000$ shown in part (d) there is little difference compared to the standard case. Only the secondary structures at large negative values of $L$ have changed.

\subsection{Dependence on the major axis of the bar}
\label{semi}

\begin{figure}
\begin{center}
\includegraphics[width=\hsize]{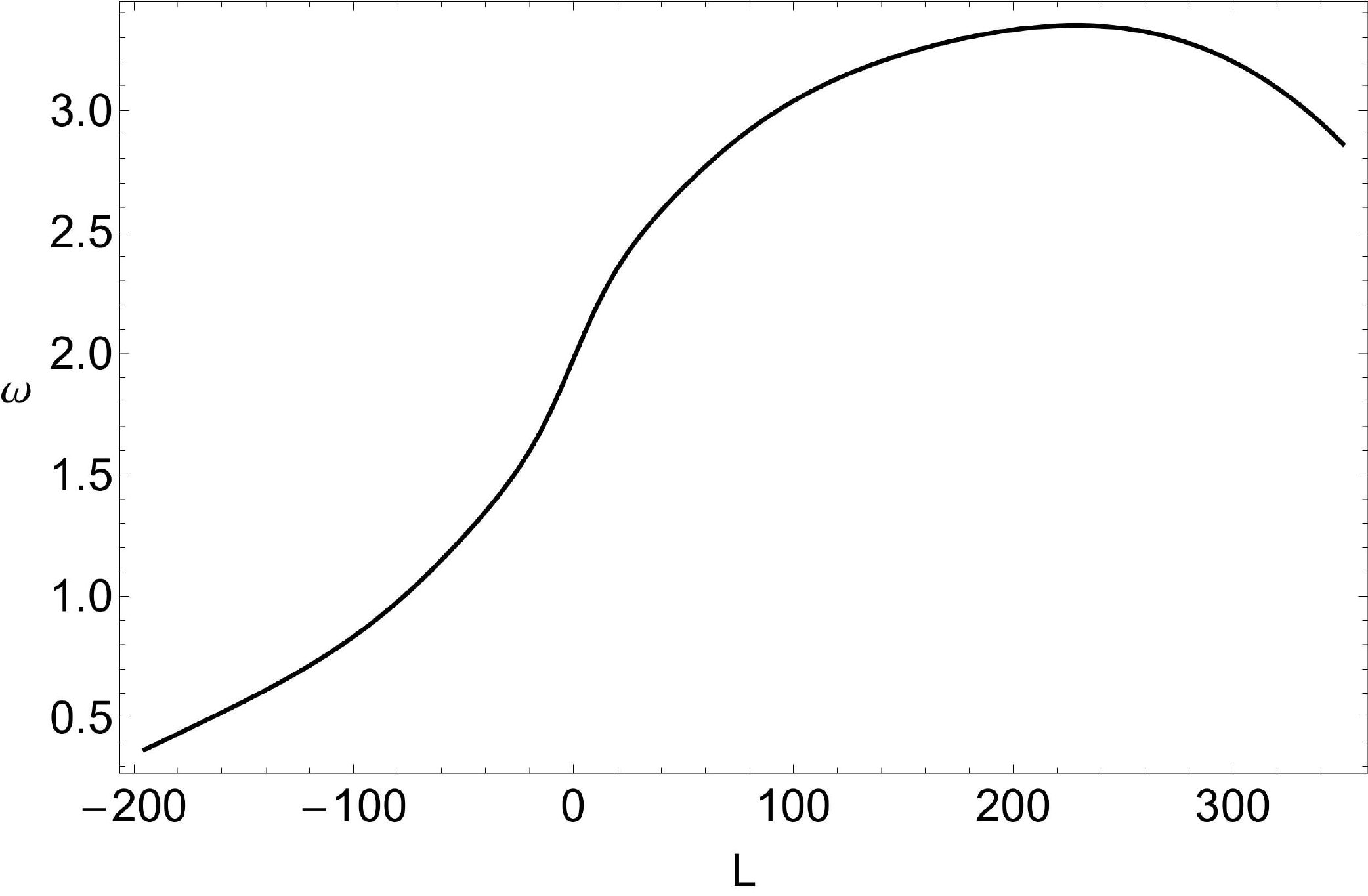}
\end{center}
\caption{The rotation angle $\omega$ as a function of the angular momentum $L$ for $a = 0$.}
\label{a0}
\end{figure}

\begin{figure*}
\centering
\resizebox{\hsize}{!}{\includegraphics{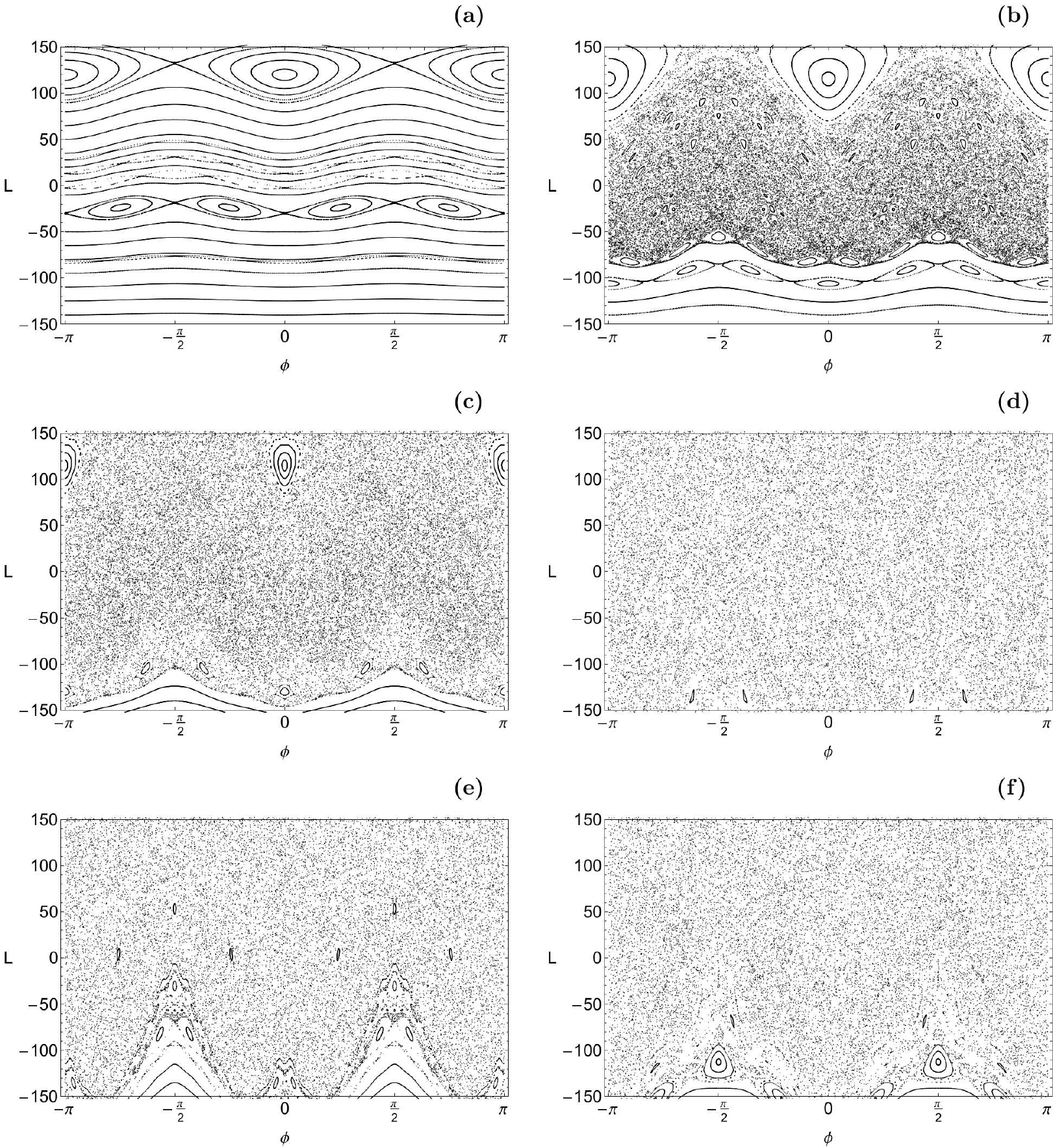}}
\caption{Examples of the perturbed map $P_r$ for various values of the major axis of the bar $a$ when $E = -900$. (a): $a = 1$; (b): $a = 2$; (c): $a = 3$; (d): $a = 5$; (e) $a = 7$; (f): $a = 9$.}
\label{ab}
\end{figure*}

\begin{figure*}
\centering
\resizebox{\hsize}{!}{\includegraphics{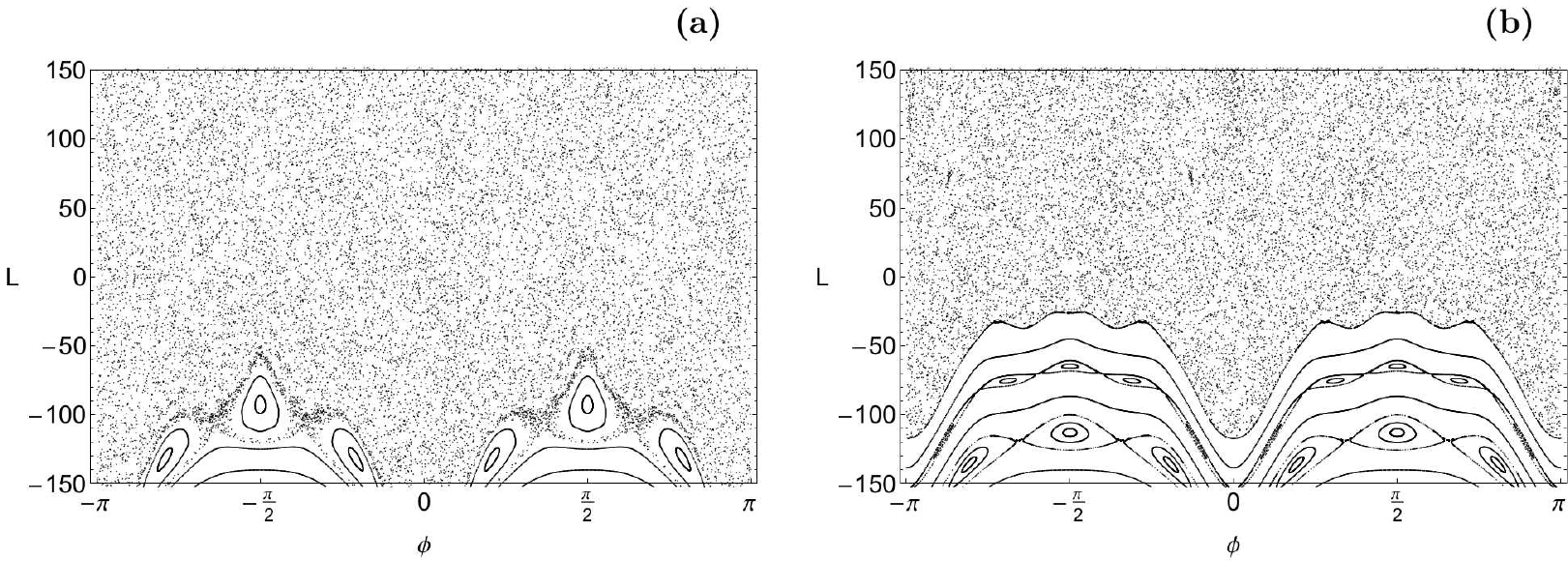}}
\caption{Examples of the perturbed map $P_r$ for various values of the scale length of the bar $c_{\rm b}$ when $E = -900$. (a-left): $c_{\rm b} = 0.1$; (b-right): $c_{\rm b} = 2.5$.}
\label{cb}
\end{figure*}

\begin{figure*}
\centering
\resizebox{\hsize}{!}{\includegraphics{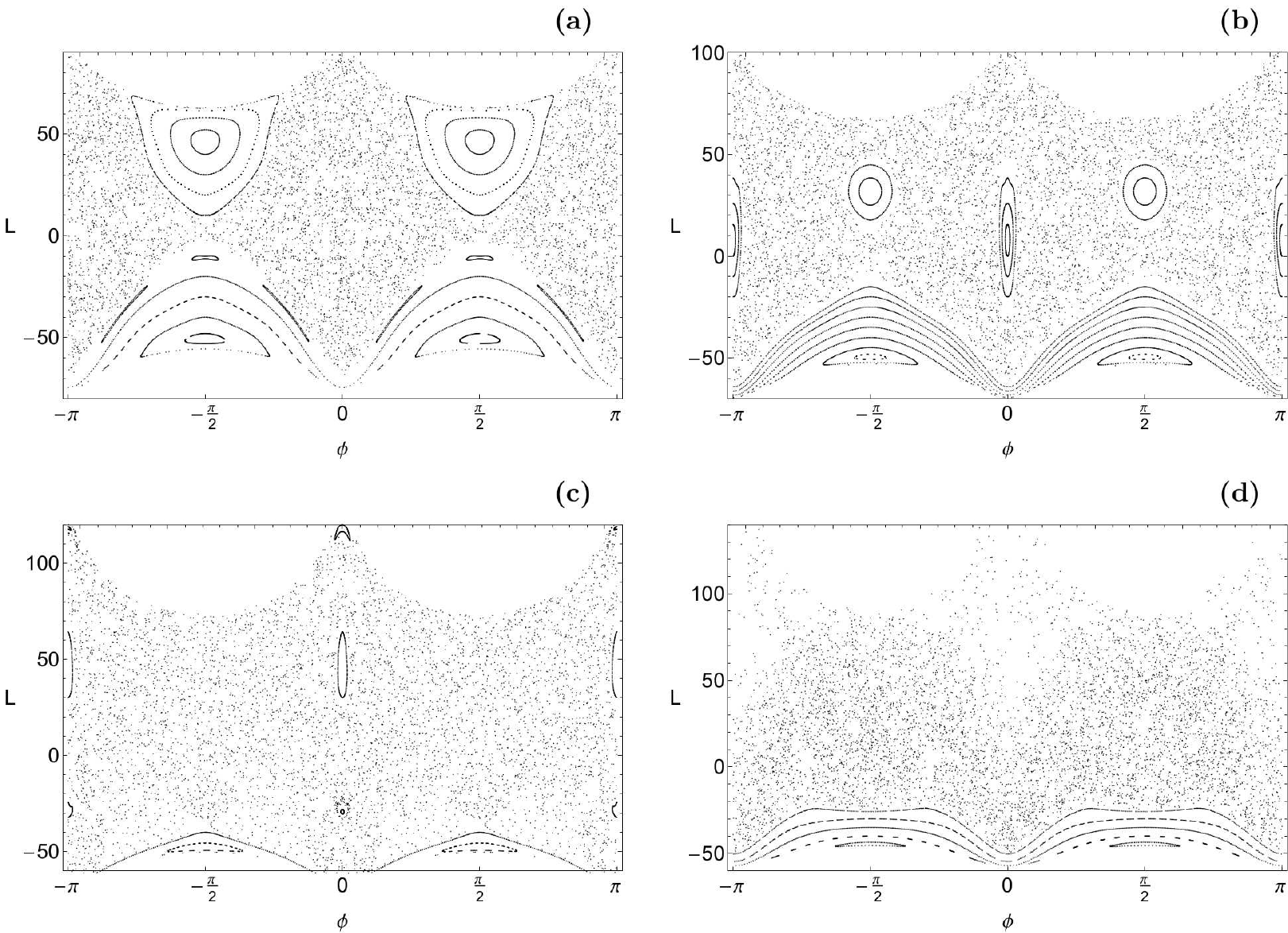}}
\caption{Examples of the perturbed map $P_r$ for various values of the rotational speed of the bar $\Omega_{\rm b}$ when $E = -2400$. (a-upper left): $\Omega_{\rm b} = 1$; (b-upper right): $\Omega_{\rm b} = 2$; (c-lower left): $\Omega_{\rm b} = 3$; (d-lower right): $\Omega_{\rm b} = 5$.}
\label{om}
\end{figure*}

\begin{figure*}
\centering
\resizebox{\hsize}{!}{\includegraphics{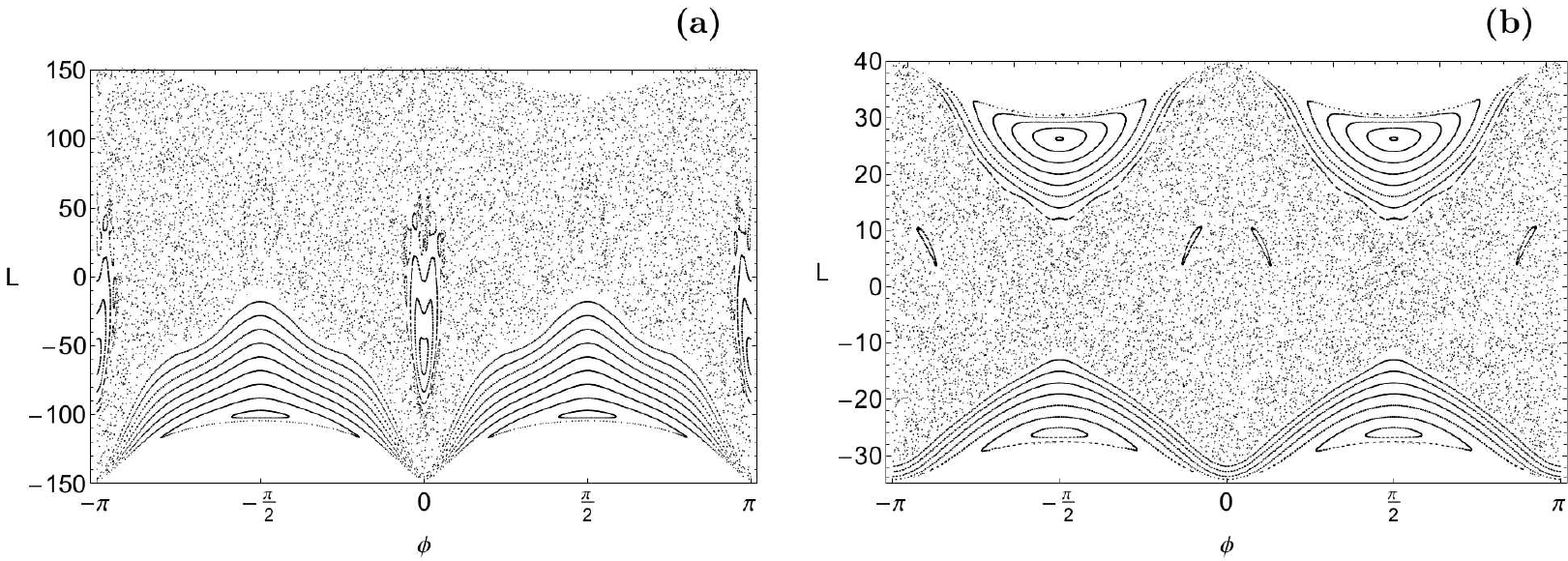}}
\caption{Examples of the perturbed map $P_r$ for various values of the total orbital energy $E$. (a-left): $E = -1650$; (b- right): $E = -3000$.}
\label{en}
\end{figure*}

For $a = 0$ the bar is rotationally symmetric and therefore we have again a case where $L$ is conserved and $P_r$ is again a pure twist map which is characterized by its twist curve. The twist curve of this limiting case is presented in Fig. \ref{a0}. The most important resonances are a 1:2 resonance at $L \approx 120$, a 2:5 resonance at $L \approx 40$, a 3:8 resonance at $L \approx 20$, a 1:3 resonance at $L \approx 0$, a 1:4 resonance at $L \approx -25$  and a 1:6 resonance at $L \approx -80$. The perturbation of this twist map under a change of the parameter $a$ is presented in Fig. \ref{ab}. For the small value $a = 1$ in part (a) we see the secondary structures expected from the twist curve of Fig. \ref{a0}. Remember that again the odd period structures of the 2:5 resonance and the 1:3 resonance come in two distinct copies. For this small value of the perturbation all fine chaos strips are still very small and appear in the plots like separatrix curves.

For $a = 2$ plotted in part (b) the $L$ interval $(-60,80)$ is already dominated by a large scale chaotic sea in which many of the previously seen secondary structures are dissolved. For $a = 3$ shown in part (c) we see small remnants of the secondary 1:2 and 1:6 islands. Otherwise the only regular structures on a visible scale in the plot are the primary KAM curves for large negative values of $L$. There is another pair of islands at very large positive values of $L \approx 270$ outside of the frame of the figure. For further increasing values of $a$ we approach the structure of the standard case without any further important changes in the large scale structure.

\subsection{Dependence on the scale length of the bar}
\label{scal}

For $c_{\rm b} \to \infty$ we approach a rotationally symmetric case which is equivalent to the case $M_{\rm b} = 0$. However, very large values of $c_{\rm b}$ do not describe realistic models of real galaxies. Therefore we do not describe this limiting case in detail. In Fig. \ref{cb} we just show the cases $c_{\rm b} = 0.1$ and $c_{\rm b} = 2.5$ to demonstrate that for realistic values of $c_{\rm b}$ there is little development of the dynamics at all under changes of $c_{\rm b}$.

\subsection{Dependence on the rotational velocity of the bar}
\label{speed}

For increasing values of $\Omega_{\rm b}$ the dynamics becomes rather rapidly unstable because of the increasing centrifugal forces. Then most orbits are scattering orbits which come from infinity and return to infinity. In this article our topic is not the chaotic scattering dynamics, we are only interested in bound motion. Therefore we describe the dependence on $\Omega_{\rm b}$ for a smaller value of the energy $E$ where still a large part of the motion is bound also for values of $\Omega_{\rm b}$ up to 5. For the numerical examples we choose the value $E = -2400$. Some examples of the corresponding map $P_r$ are plotted in Fig. \ref{om}. For $\Omega_{\rm b}$ converging to zero the map becomes upside
down symmetric in $L$ because for $\Omega_{\rm b} = 0$ the two orientations of angular motion become equivalent and correspondingly the sign of $L$ becomes irrelevant. For $\Omega_{\rm b}$ small and $L$ not too close to zero the motion becomes stable for $\phi$ values around $\pm \pi/2$. That is, motion around the origin and mainly perpendicular to the bar becomes dynamically stable.

In part (a) of the figure for $\Omega_{\rm b} = 1$ we see two distinct island chains of period 2 with centers at $\phi = \pm \pi/2$, one at positive and one at large negative values of $L$. Increasing values of $\Omega_{\rm b}$ break the symmetry between the two rotational orientations, i.e. between the two signs of $L$. The motion with positive values of $L$ becomes
unstable soon whereas the motion with large negative values of $L$ remains stable up to very large values of $\Omega_{\rm b}$. Note that for intermediate values of $\Omega_{\rm b}$ (see parts (b) and (c) of the figure) motion with small values of
$L$ and along the direction of the bar becomes stable. See the islands at $\phi = 0$ and $\phi = \pi$. The coming and going of these islands indicates that motion along the bar never becomes very unstable and that this type of motion can play the role of an important organization center of the dynamics also when it is slightly unstable.

In part (d) of the figure for $\Omega_{\rm b} = 5$ almost all the orbits of the large scale chaotic sea are in reality scattering orbits which go to infinity in the long run. For typical initial conditions in this chaotic sea the orbit first increases its value of $L$ to values above 500 and then the value of $R$ becomes large monotonically, i.e. the orbit reaches the asymptotic region and does no longer produce any intersections with the surface of section of the Poincar\'{e} map. This transition to the asymptotic behaviour is not the topic of the present article.

\subsection{Dependence on the orbital energy}
\label{ener}

We have already seen a lot of maps for $E = -900$ and in the $\Omega_{\rm b}$ series also some for $E=-2400$. Finally let us see two more examples for different values of $E$. Fig. \ref{en}(a-b) present the examples for $E=-1650$ and $E=-3000$, respectively, otherwise the plots are for the standard values of the bar parameters. In Fig. \ref{en} we see again the stable islands at $\phi = 0$ and $\phi = \pi$ which also exist for some other parameter combinations. And for $E = -3000$ we see the large scale islands of period 2 at $\phi = \pm \pi/2$ and for both maximal positive and negative values of $L$.

The minimal possible value of the energy for the standard bar parameters is $E = E_{\rm min} = -4854$. In the limit $E \to E_{\rm min}$ the motion converges to completely integrable harmonic motion even in 3-dof. This is the last integrable limit case encountered in our study.

Certainly the most persistent type of stable motion in $S_z$ is the motion at the maximal negative values of $L$ and in particular for angles in the neighbourhood of $\phi = \pm \pi/2$. We did not find any combination of reasonable realistic parameter values which makes this motion unstable. It is motion in negative orientation around the center where the maximal values of $R$ occur in the direction perpendicular to the bar. In comparison, motion mainly in direction of the bar is stable for limited parameter regions only. As Fig. \ref{ab} shows, this last mentioned motion grows out of the 1:2 resonance of the integrable limit case $a = 0$ and it becomes unstable for $a \approx 4.7$.

\subsection{The x1 orbital family}
\label{x1o}

As seen in Fig. \ref{ab} for small values of $a$ the 1:2 resonance in the Poincar\'{e} map gives the largest secondary KAM islands. In this sense the periodic orbit belonging to the central periodic point of these islands plays a prominent role in the dynamics. From the Poincar\'{e} maps we observe that this orbit stays stable until $a \approx 4.7$. And we expect that it still serves as an important organizing center of the dynamics also for larger values of $a$. Therefore in this subsection we study this periodic orbit in more detail.

\begin{figure*}
\centering
\resizebox{0.80\hsize}{!}{\includegraphics{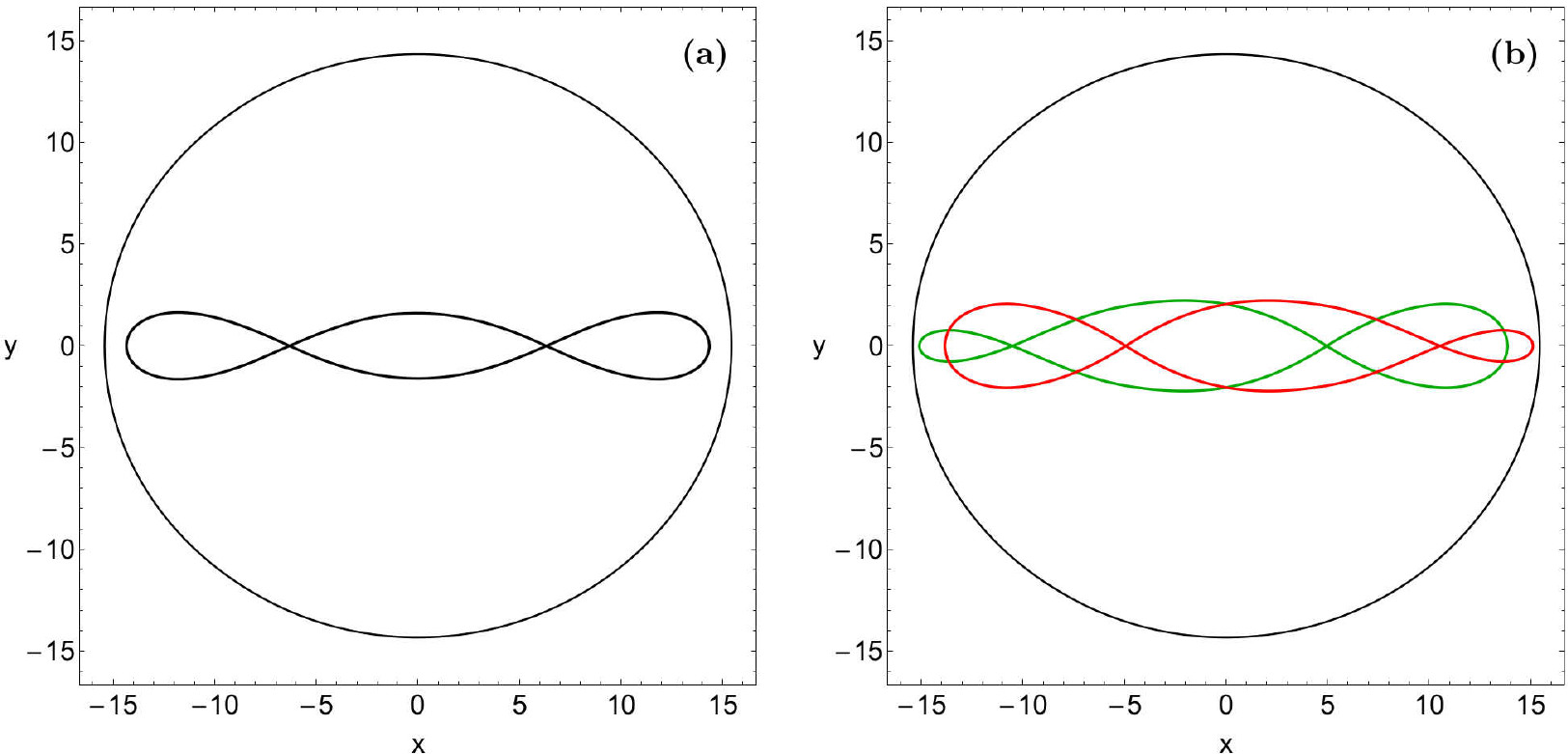}}
\caption{(a-left): An unstable x1 1:3 periodic orbit for $a = 7$ when $E = -900$; (b-right): The two left-right asymmetric x1 1:3 periodic orbits for $a = 7$ when $E = -900$. Note that one orbit is the mirror image of the other one.}
\label{x1}
\end{figure*}

In Fig. \ref{x1}a we show the orbit in position space for the value $a = 7$. It is evident that it is an orbit moving mainly in the direction of the bar ($x$-axis) making at the same time smaller oscillations perpendicular to the bar directions(i.e., along the $y$-axis). Between these longitudinal and transverse oscillations in the bar potential a 1:3 resonant coupling is evident. Orbits that are elongated parallel to the bar within corotation are known as x1 orbits and in this sense our main orbit is a 1:3 resonant x1 orbit. Along one complete revolution this orbits runs through 2 maxima of $R$ and therefore it appears as period 2 point in our Poincar\'{e} sections with the intersection condition that $R$ is maximal.

\begin{figure*}
\centering
\resizebox{0.80\hsize}{!}{\includegraphics{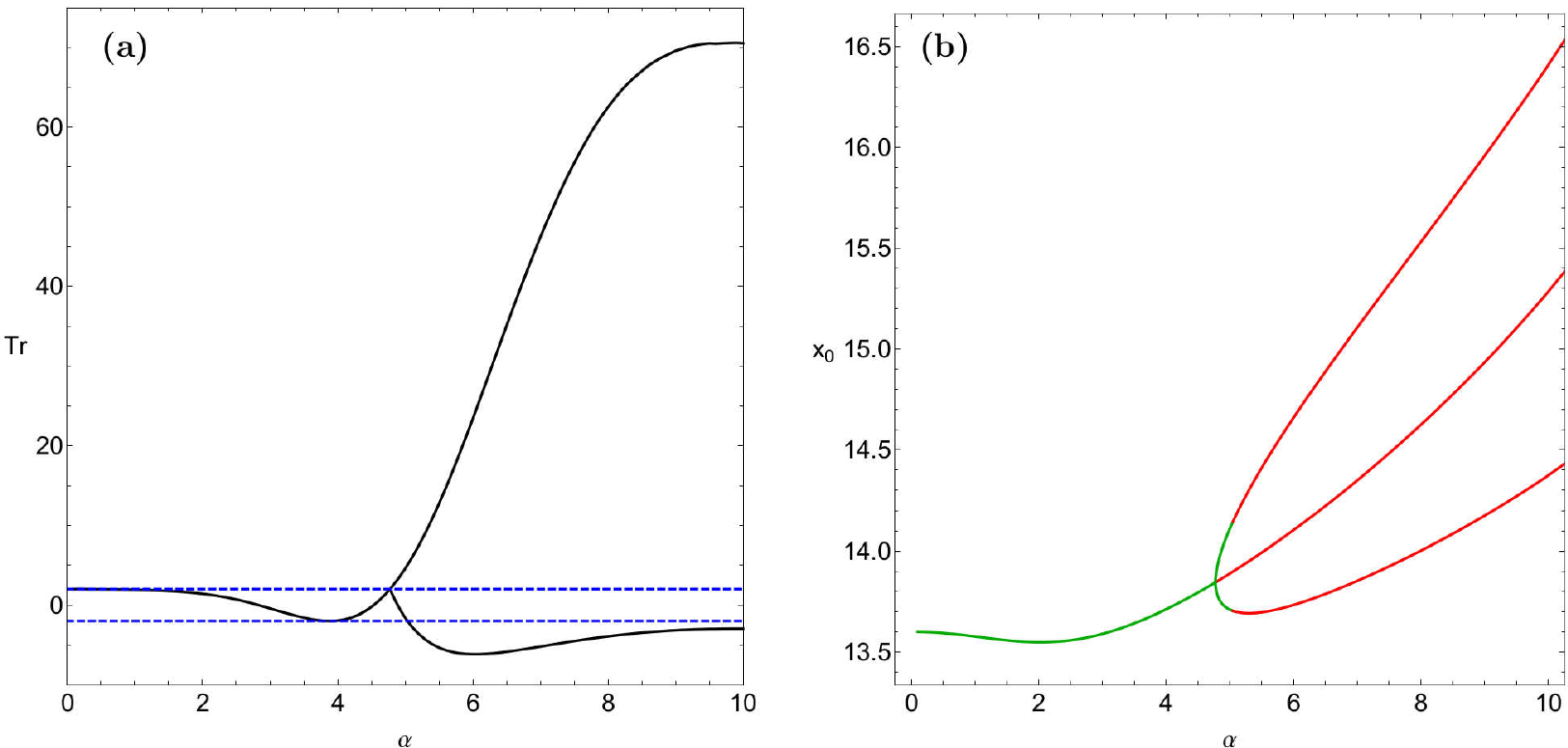}}
\caption{(a-left): Evolution of the trace of the monodromy matrix of the x1 1:3 family as a function of the major axis of the bar $a$ when $E = -900$. The blue horizontal dashed lines at --2 and +2 delimit the range of the trace for which the periodic orbits are stable; (b-right): The characteristic curve of the x1 1:3 family.}
\label{fpos}
\end{figure*}

In the next Fig. \ref{fpos}a we present graphically the stability type of this orbit, we plot the trace $Tr$ of its monodromy matrix as function of $a$ as the black curve. Remember that a periodic orbit is elliptic if $-2 < Tr <2$, it is normal parabolic if $Tr = 2$, normal hyperbolic if $Tr > 2$, inverse parabolic if $Tr = -2$ and inverse hyperbolic if $Tr < -2$. The eigenvalues $\lambda$ of the monodromy matrix are obtained from the trace as
\begin{equation}
\lambda = Tr/2 \pm \sqrt{(Tr/2)^2 - det}
\end{equation}
And because the monodromy matrix is symplectic we always have $det = 1$ for its determinant. Because of the importance of the values +2 and -2 for the trace two horizontal dashed lines (blue in the colour version) at these values are included in the figure.

As we have mentioned before, for $a = 0$ we have a pure twist map which is globally parabolic and therefore the curve in Fig. \ref{fpos}a must start with the value $Tr = 2$ at $a=0$. With increasing value of $a$ the eigenvalues run along the complex unit circle and when the phase of the eigenvalues reaches $\pi$ then the trace reaches the value -2. Because of discrete symmetry the orbit does not become inverse hyperbolic at this point. It stays elliptic and the eigenvalue continues to run along the unit circle. For $a \approx 4.77$ the eigenvalue reaches +1, the trace reaches +2, and here the orbit becomes normal hyperbolic in a pitchfork bifurcation. The trace curve of the original 1:3 x1 orbit passes through this bifurcation as smooth curve. At the moment of the change of stability the original orbit splits off two descendants where each one breaks the left-right symmetry. But one of them is the mirror image of the other one. Fig. \ref{x1}b shows these two periodic orbits in position space for $a = 7$.

Fig. \ref{fpos}b presents a bifurcation diagram of the pitchfork bifurcation. Here the maximal value of the $x$ coordinate of each one of the three orbits is plotted as function of $a$. The curve for $a < 4.7$ and the middle curve for $a > 4.7$ belong to the original symmetric 1:3 x1 orbit. The two outer branches for $a > 4.7$ belong to the two orbits split off in the pitchfork bifurcation. As long as the corresponding orbit is stable the curve is green and when the orbit is unstable the curve in Fig. \ref{fpos}b is red.

Also included in Fig. \ref{fpos}a for $a > 4.77$ is the trace of the monodromy matrix of the asymmetric orbits split off in the pitchfork bifurcation. It is the lower curve in the figure. The split off orbits are born with $Tr = 2$ and are stable in a small $a$ interval. At $a \approx 5.04$ the trace passes through the value -2 and the orbits become inverse hyperbolic in a period doubling bifurcation. The orbits do not become very unstable for increasing $a$. After a first increase the instability even returns to rather moderate values and around $a = 10$ the orbits are almost stable. Interestingly also the instability of the original 1:3 x1 orbit saturates for these values of $a$. All these orbits become very unstable only for very large values of $a$ outside of the range where the model makes physical sense. For extreme values of $a$ the split off orbits change from 1:3 x1 type to 1:2 x1 type, i.e. then they make 2 transverse oscillations only for each longitudinal oscillation along the bar.

In total we find the following situation for $a \in (6,10)$: There is a whole chaotic braid of 1:3 x1 orbits, the original left-right symmetric one, the two ones split of in the pitchfork bifurcations breaking the left-right symmetry, and the infinity created in the doubling cascades of the split off orbits. At least some of them have rather moderate instability which implies that also the instability of the whole braid is moderate. And this in turn implies that this chaotic braid serves as an important organizing center for the whole dynamics.

\section{NUMERICAL RESULTS FOR THE 3-DOF SYSTEM}
\label{numres3}

\begin{figure}
\begin{center}
\includegraphics[width=\hsize]{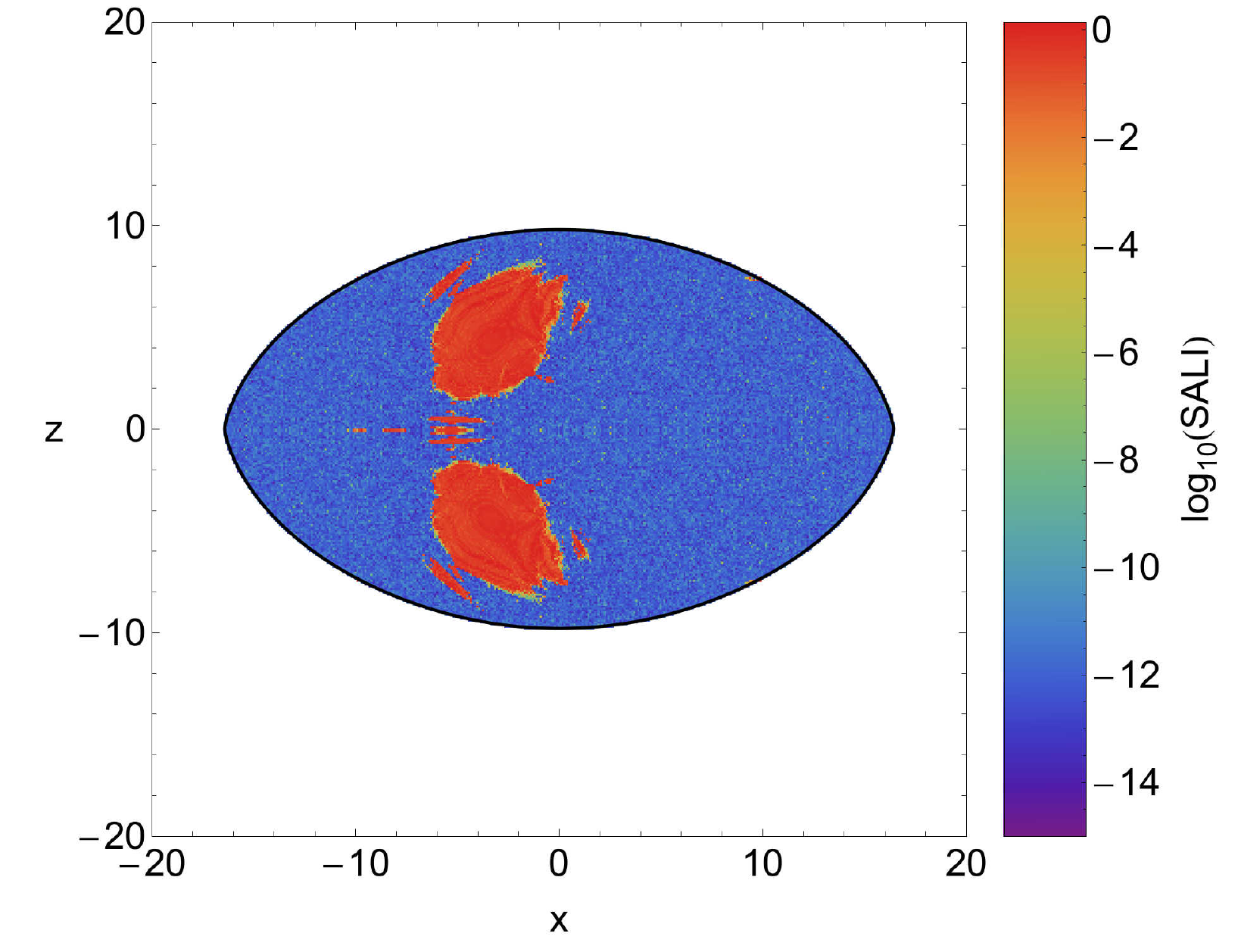}
\end{center}
\caption{Regions of different values of SALI in a dense grid of initial conditions on the $(x,z)$-plane for the standard model.}
\label{sm31}
\end{figure}

\begin{figure}
\begin{center}
\includegraphics[width=\hsize]{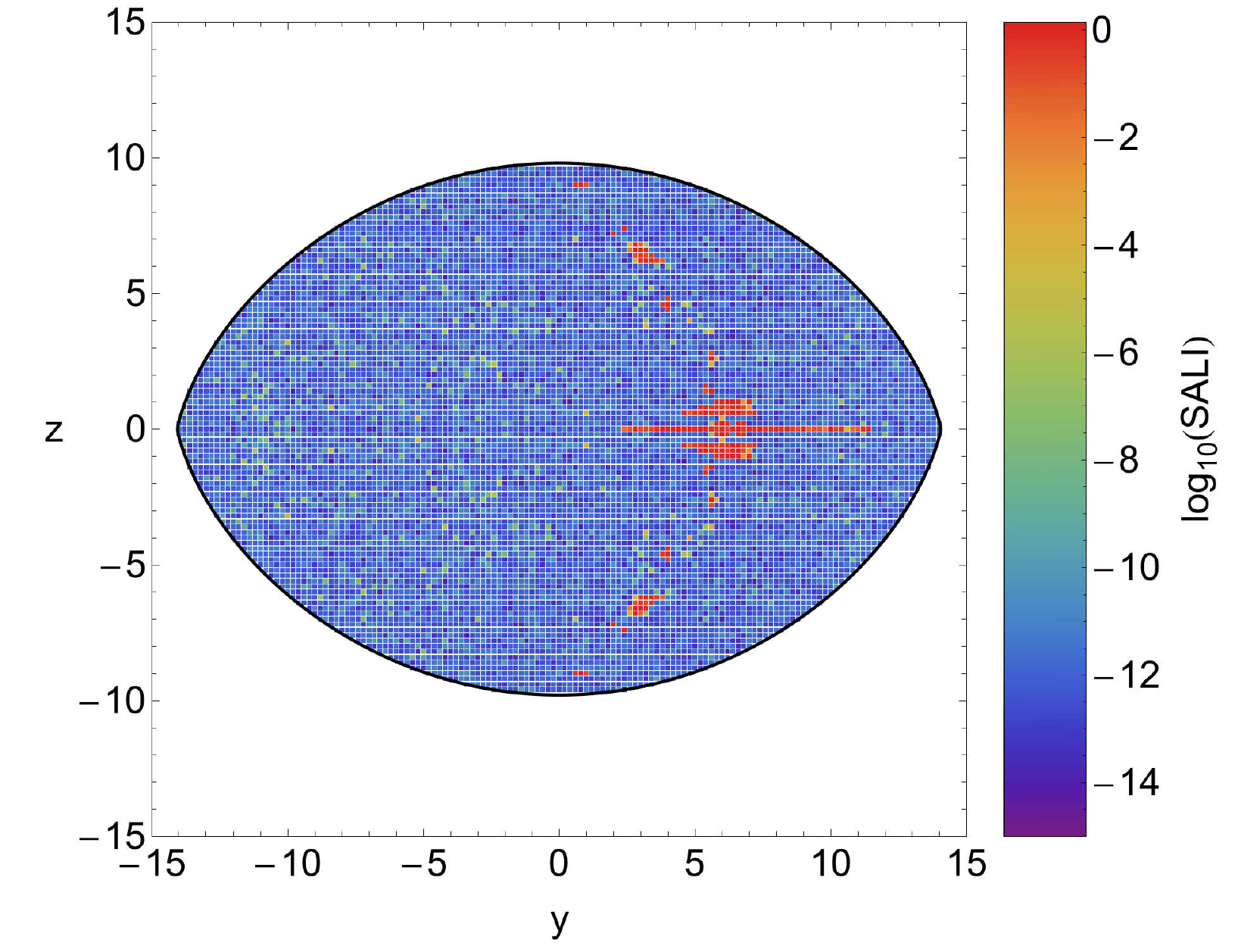}
\end{center}
\caption{Regions of different values of SALI in a dense grid of initial conditions on the $(y,z)$-plane for the standard model.}
\label{sm32}
\end{figure}

\begin{figure*}
\centering
\resizebox{\hsize}{!}{\includegraphics{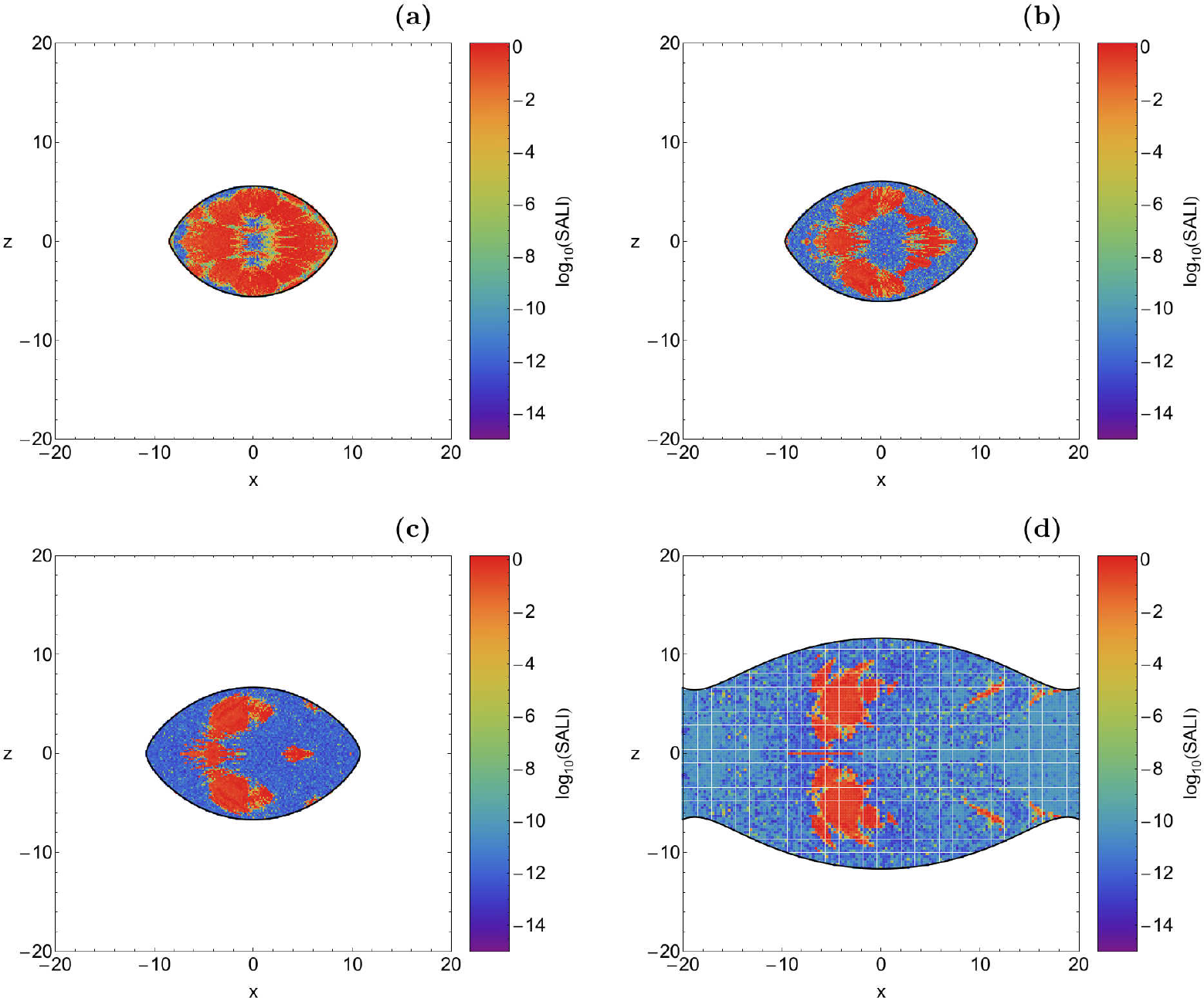}}
\caption{Regions of different values of SALI in a dense grid of initial conditions on the $(x,z)$-plane when $E = -900$. (a-upper left): $M_{\rm b} = 100$; (b-upper right): $M_{\rm b} = 500$; (c-lower left): $M_{\rm b} = 1000$; (d-lower right): $M_{\rm b} = 5000$.}
\label{mb3}
\end{figure*}

\begin{figure*}
\centering
\resizebox{\hsize}{!}{\includegraphics{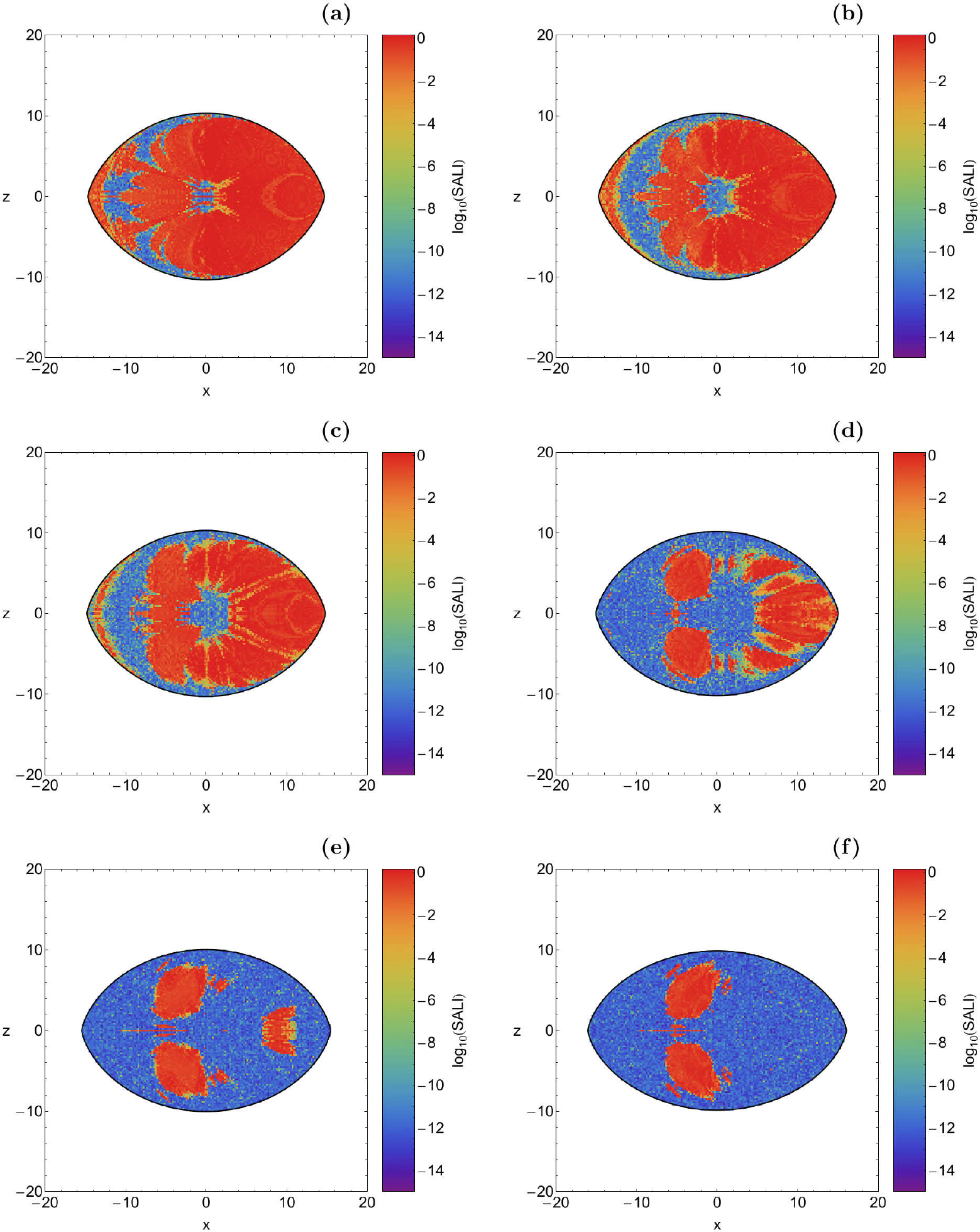}}
\caption{Regions of different values of SALI in a dense grid of initial conditions on the $(x,z)$-plane when $E = -900$. (a): $a = 1$; (b): $a = 2$; (c): $a = 3$; (d): $a = 5$; (e) $a = 7$; (f): $a = 9$.}
\label{ab3}
\end{figure*}

\begin{figure*}
\centering
\resizebox{\hsize}{!}{\includegraphics{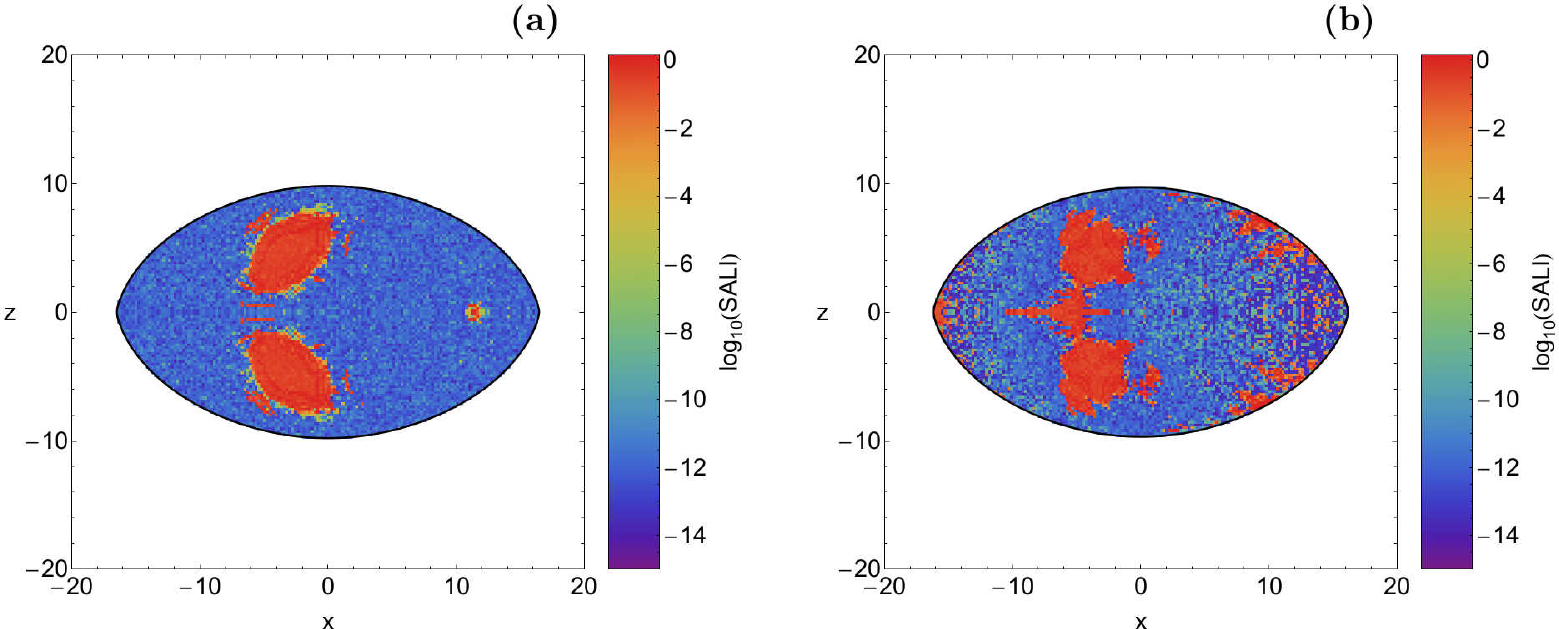}}
\caption{Regions of different values of SALI in a dense grid of initial conditions on the $(x,z)$-plane when $E = -900$. (a-left): $c_{\rm b} = 0.1$; (b-right): $c_{\rm b} = 2.5$.}
\label{cb3}
\end{figure*}

\begin{figure*}
\centering
\resizebox{\hsize}{!}{\includegraphics{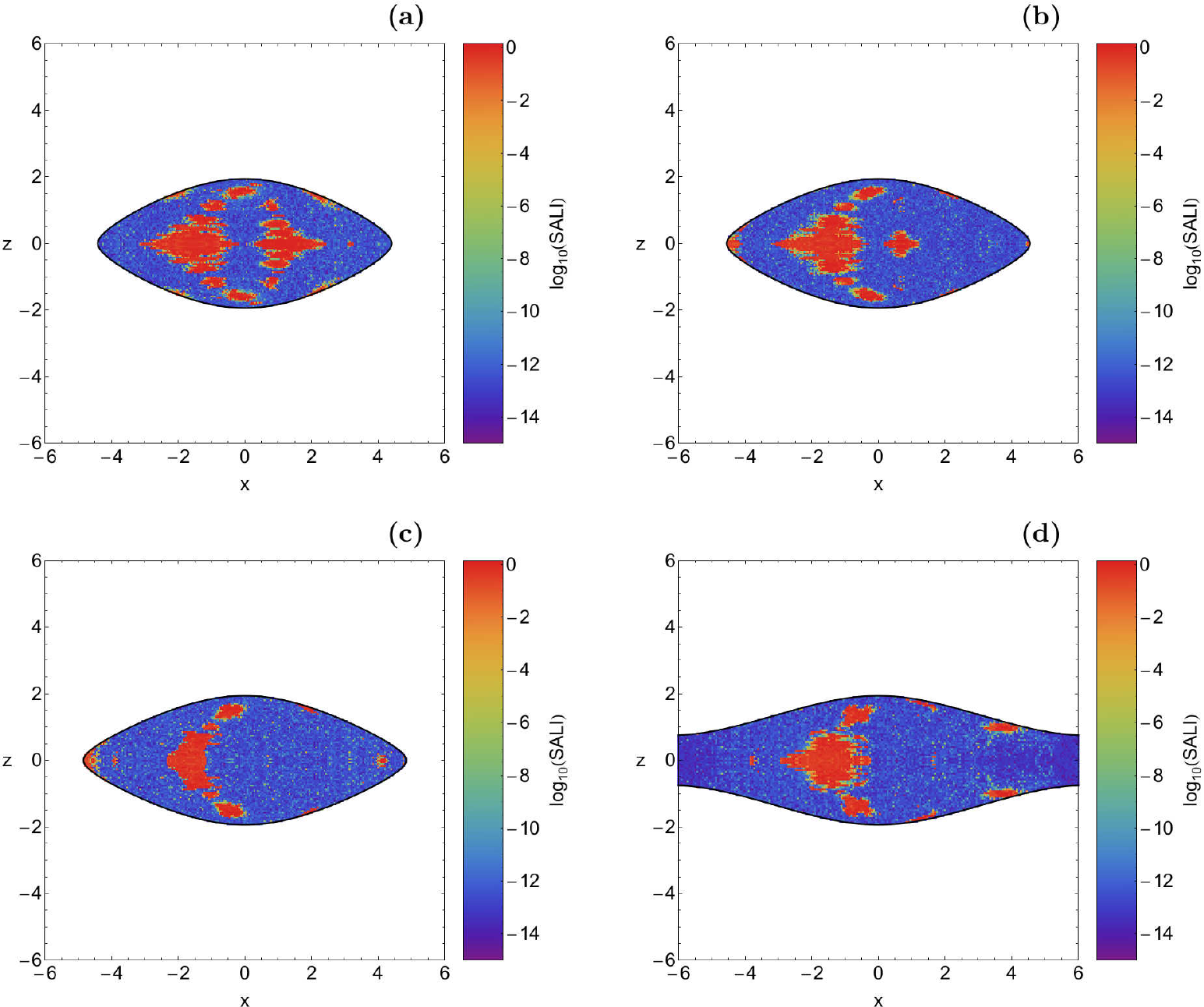}}
\caption{Regions of different values of SALI in a dense grid of initial conditions on the $(x,z)$-plane when $E = -2400$. (a-upper left): $\Omega_{\rm b} = 1$; (b-upper right): $\Omega_{\rm b} = 2$; (c-lower left): $\Omega_{\rm b} = 3$; (d-lower right): $\Omega_{\rm b} = 5$.}
\label{om3}
\end{figure*}

\begin{figure*}
\centering
\resizebox{\hsize}{!}{\includegraphics{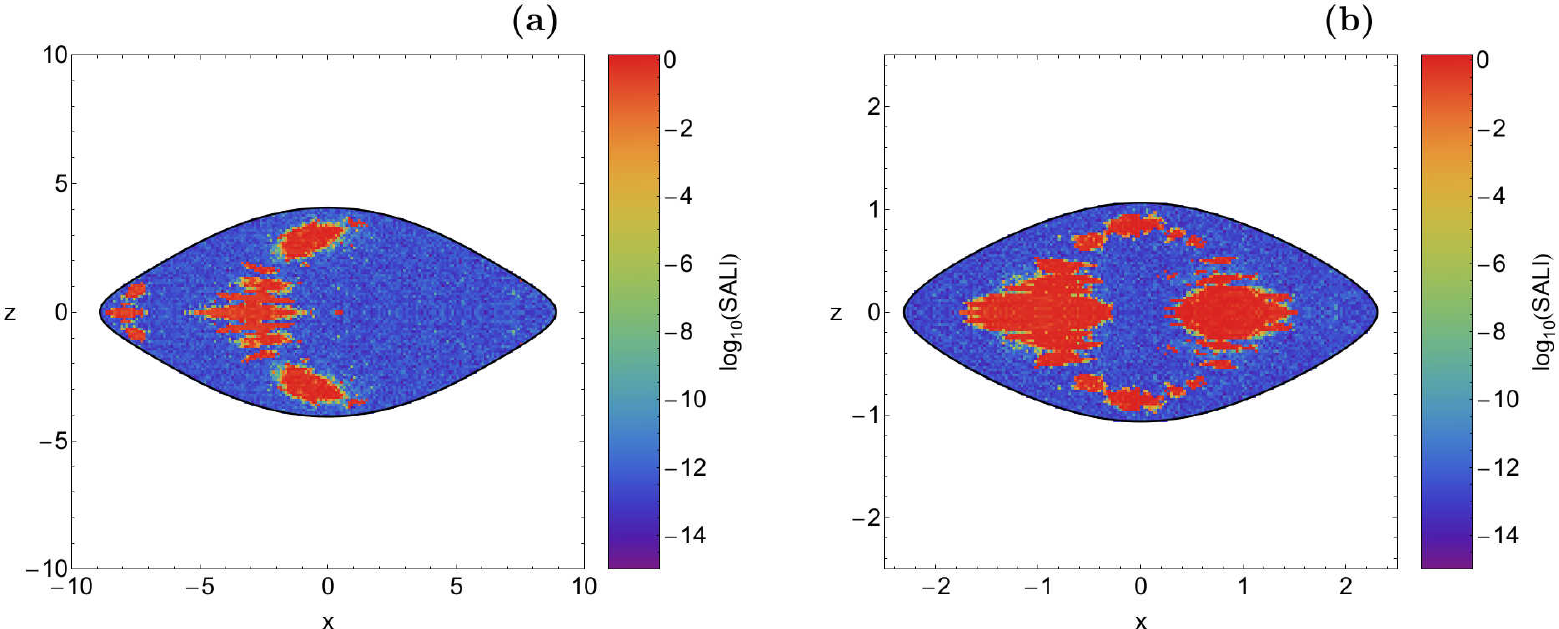}}
\caption{Regions of different values of SALI in a dense grid of initial conditions on the $(x,z)$-plane. (a-left): $E = -1650$; (b- right): $E = -3000$.}
\label{en3}
\end{figure*}

\begin{figure*}
\centering
\resizebox{\hsize}{!}{\includegraphics{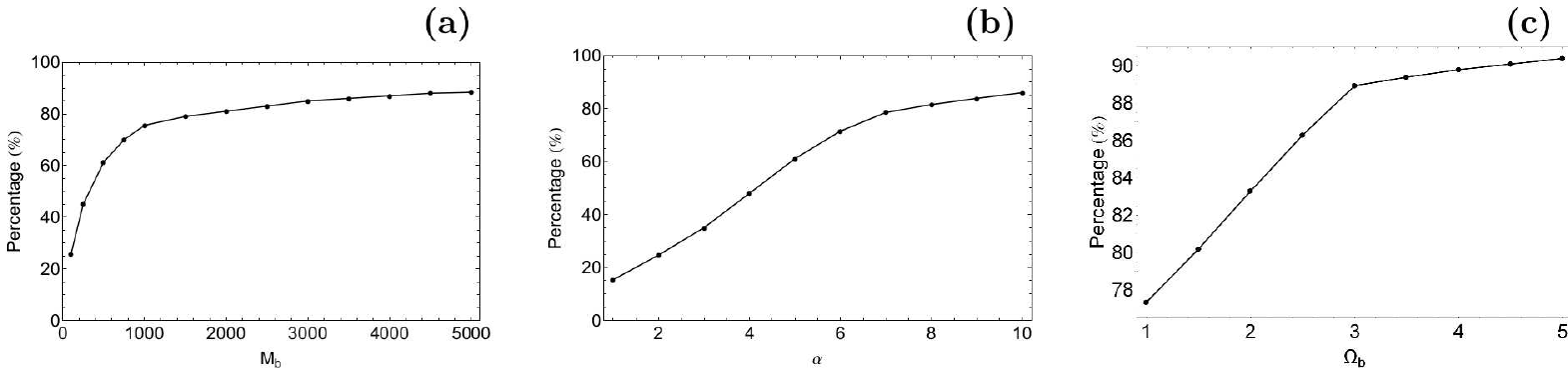}}
\caption{Evolution of the relative fraction of chaotic orbits found in the SALI plots on the $(x,z)$-planes as function of (a): $M_{\rm b}$, (b): $a$ and (c): $\Omega_{\rm b}$.}
\label{p3D}
\end{figure*}

\begin{figure}
\begin{center}
\includegraphics[width=\hsize]{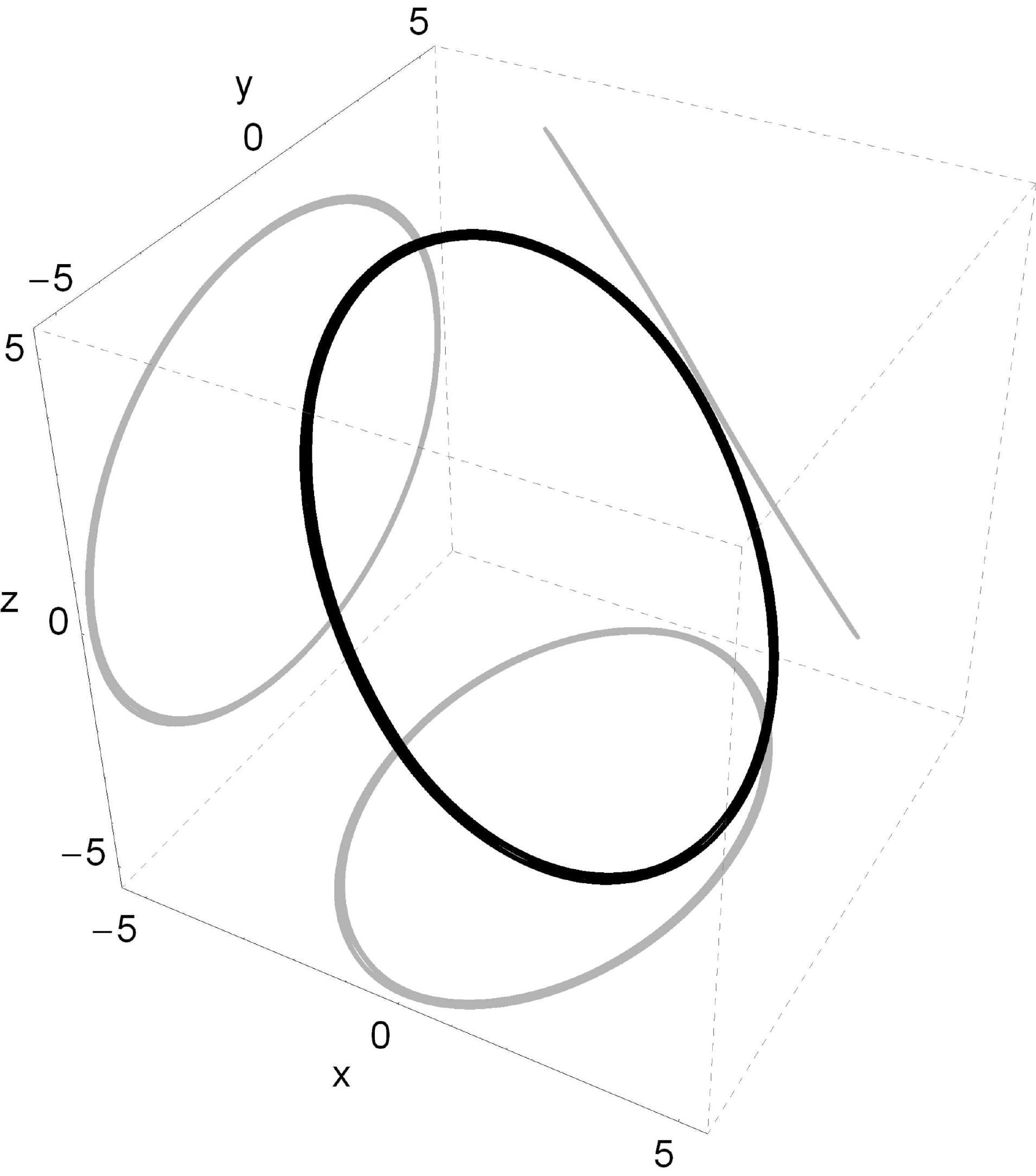}
\end{center}
\caption{A characteristic example of an inclined 3D loop orbit along with its projections on the three primary planes.}
\label{orb3}
\end{figure}

A simple qualitative way for distinguishing between ordered and chaotic motion in a Hamiltonian system is by plotting the successive intersections of the orbits using a Poincar\'{e} surface of section (PSS). This method has been extensively applied to 2-dof models, as in these systems the PSS is a two-dimensional plane (see Section \ref{numres2}). In 3-dof systems, however, the PSS is four-dimensional and thus the behaviour of the orbits cannot be easily visualized.

One way to solve this problem is to choose a 2 dimensional plane in the phase space, cover it with a sufficiently fine grid of points, use these points as initial conditions for orbits and check for each one of these orbits whether its motion is regular or chaotic, following the method used in \citet{MA11,ZC13,Z14}. In this way, we are able to identify again regions of order and chaos, which may be visualized, because we choose as domains only planes of dimension 2. For most of our SALI plots we use as domain the $(x,z)$-plane, choose for each orbit an initial point $(x_0, z_0)$ in this plane, take 3 other initial conditions as $y_0 = p_{x0} = p_{z0} = 0$ and obtain the last initial condition $p_{y0}$ from the value of the energy according to Eq. (\ref{ham}), where we use the positive branch of the solution. Thus, we are able to construct again a two-dimensional plot depicting the $(x,z)$-plane. All the initial conditions of the three-dimensional orbits lie inside the limiting curve defined by
\begin{equation}
f(x,z) = \Phi_{eff}(x,y = 0,z) = E.
\label{zvc}
\end{equation}
In a few examples we also show SALI plots on the $(y,z)$-plane as domain. They can be understood by exchanging the roles of $x$ and $y$ in the above mentioned construction.

For distinguishing between order and chaos in our models we use the SALI method \citep{S01}. We chose for each value of the free parameter of the effective potential, a dense uniform grid of $10^{5}$ initial conditions in the $(x,z)$-plane, regularly distributed in the area allowed by the value of the energy $E$. For each initial condition, we integrated the equations of motion (\ref{eqmot}) as well as the variational equations (\ref{vareq}) with a double precision Bulirsch-Stoer algorithm \citep{PTVF92}. In all cases, the value of the Jacobi integral (Eq. (\ref{ham})) was conserved better than one part in $10^{-11}$, although for most orbits, it was better than one part in $10^{-12}$.

All initial conditions of orbits are numerically integrated for $10^{3}$ time units which correspond to about $10^{11}$ yr or in other words to about 10 Hubble times. This vast time of numerical integration is justified due to the presence of the sticky orbits. Therefore, if the integration time is too short, any chaos indicator will misclassify sticky orbits as regular ones. In our work we decided to integrate all orbits for a time interval of $10^{3}$ time units in order to correctly classify sticky orbits with sticky periods of at least of 10 Hubble times. At this point, it should be clarified that sticky orbits with sticky periods larger than $10^{3}$ time units will be counted as regular ones, since such extremely high sticky periods are completely out of scope of this research.

Now we use the SALI plots to observe the parameter dependence of the dynamics of the full 3-dof system. To each Poincar\'e map from the previous section we show the corresponding SALI plot. We begin with the standard model in Figs. \ref{sm31} and \ref{sm32}, where Fig. \ref{sm31} is the SALI plot in the $(x,z)$-plane and Fig. \ref{sm32} is the SALI plot in the $(y,z)$-plane. The corresponding Poincar\'e plot has already been presented in Fig. \ref{sm}a.

Let us explain the connection between the Poincar\'e plot in the $(\phi,L)$-plane and the SALI plot in the $(x,z)$-plane. The line $z = 0$ in the SALI plot belongs to the invariant subset $S_z$. The part for $x$ positive corresponds to $\phi = 0$ and the part for $x$ negative corresponds to $\phi = \pi$. For all points in the domain of the SALI plot we have $y = 0$ and also $p_x = 0$, therefore automatically we have $p_R = 0$. Remember that the intersection condition of the Poincar\'e map is $p_R = 0$ with negative intersection orientation. i.e. $p_R$ has to change from positive to negative, i.e. $R$ has to run through a relative maximum.

We need to understand which points in the SALI plot corresponds to maxima of $R$ and which ones to minima. When an orbit starts perpendicular to the SALI plane and with small value of the coordinate $x$ then there is very little radial acceleration and the radial distance increases. In contrast, when the orbit starts with a large absolute value of $x$, then there is a strong acceleration in negative radial direction, the orbit starts with a large value of inward curvature and the initial value of $R$ is a maximum. In this sense approximately the outer half on both sides of the domain of the SALI plots corresponds to negative orientation with respect to the intersection condition $p_R = 0$ and the inner half on both sides corresponds to positive orientation of this intersection. It is also clear that points with positive $x$ correspond to positive values of $L$ and points with negative values of $x$ correspond to negative values of $L$

Accordingly, in total we have the following correspondence: Points on the outer half of the line segment $z = 0, x > 0$ in the SALI plot can be identified with points in the Poincar\'e plot along the line $\phi = 0$ and positive values of $L$ and points on the outer half of the line segment $z = 0, x < 0$ in the SALI plot can be identified with points in the Poincar\'e plot along the line $\phi = \pi$ and negative values of $L$. Analogous considerations hold for SALI plots in the $(y,z)$-plane where the corresponding values of $\phi$ are $\pm \pi/2$. Here we have only to remember that orbits start with positive values of $p_x$ and therefore positive values of $y$ in the SALI plot correspond to negative values of $L$ and negative values of $y$ in the SALI plot correspond to positive values of $L$. The outermost values of either $x$ or $y$ always correspond to $L = 0$.

Now let us check this correspondence for the plots of the standard model presented in Figs. \ref{sm}(a-b), \ref{sm31} and \ref{sm32}. In the Poincar\'e plot of Fig. \ref{sm}a we see large scale regular structures for $\phi = \pm \pi/2$ and negative values of $L$ only. Accordingly in Fig. \ref{sm31} we see no large scale regular structures at all for large absolute values of $x$ and for $z = 0$ and in Fig. \ref{sm32} we see a small strip of regular behaviour along the line $z = 0$ for positive values of $y$.

In the next figures we show SALI plots in the $(x,z)$-plane for all cases for which we have given Poincar\'e plots in the invariant plane $S_z$ in section \ref{numres2}. Fig. \ref{mb3} shows the plots in dependence on $M_{\rm b}$, Fig. \ref{ab3} contains the plots in dependence on $a$, Fig. \ref{cb3} shows the plots in dependence on $c_b$, Fig. \ref{om3} gives the plots for various values of $\Omega_{\rm b}$ and finally Fig. \ref{en3} shows the plots for two values of the total orbital energy $E$. In all cases we can confirm the correspondence between SALI plots and Poincar\'e plots which we have explained above for the standard model. In the Figs. \ref{p3D}(a-c) we show the relative fraction of chaotic orbits found in the SALI plots as function of $M_{\rm b}$, $a$ and $\Omega_{\rm b}$, respectively. Looking carefully at Fig. \ref{p3D}(a-c) it becomes evident that the percentage of the chaotic orbits in the $(x,z)$-plane increases in the following three cases: (i) when the bar becomes more massive, (ii) when the major semi-axis of the bar increases and (iii) when the rotational velocity of the bar increases. In all three cases, for large enough values of the parameters ($M_{\rm b} > 3500$, $a > 9$, $\Omega_{\rm b} > 2.3$) initial conditions of chaotic orbits dominate covering more than 85\% of the $(x,z)$-plane.

The great value of all these SALI plots lies in the following: By the comparison with the Poincar\'e plots we obtain the information to which extent the regular structures found in the Poincar\'e plots are stable against perturbations in $z$ direction. For the example of the standard model a comparison between Figs. \ref{sm}a and \ref{sm32} shows that the regular structure at $\phi= \pm \pi/2$ and negative values of $L$ in the Poincar\'e plot corresponds to a rather narrow stable strip around $z = 0$ and large values of $y$ in the SALI plot of Fig. \ref{sm32}. Accordingly this stable structure is rather sensitive against out of plane perturbations and can be destroyed by moderate out of plane perturbations. In contrast the stable structure along the line $z = 0$ in SALI plot of Fig. \ref{en3}b, the case for $E = -3000$, is very wide in $z$ direction and therefore we conclude that the stable structure seen in the Poincar\'e plot of Fig. \ref{en}b is rather stable against out of plane perturbations. Analogous considerations hold for all other parameter cases.

\begin{figure*}
\centering
\resizebox{0.8\hsize}{!}{\includegraphics{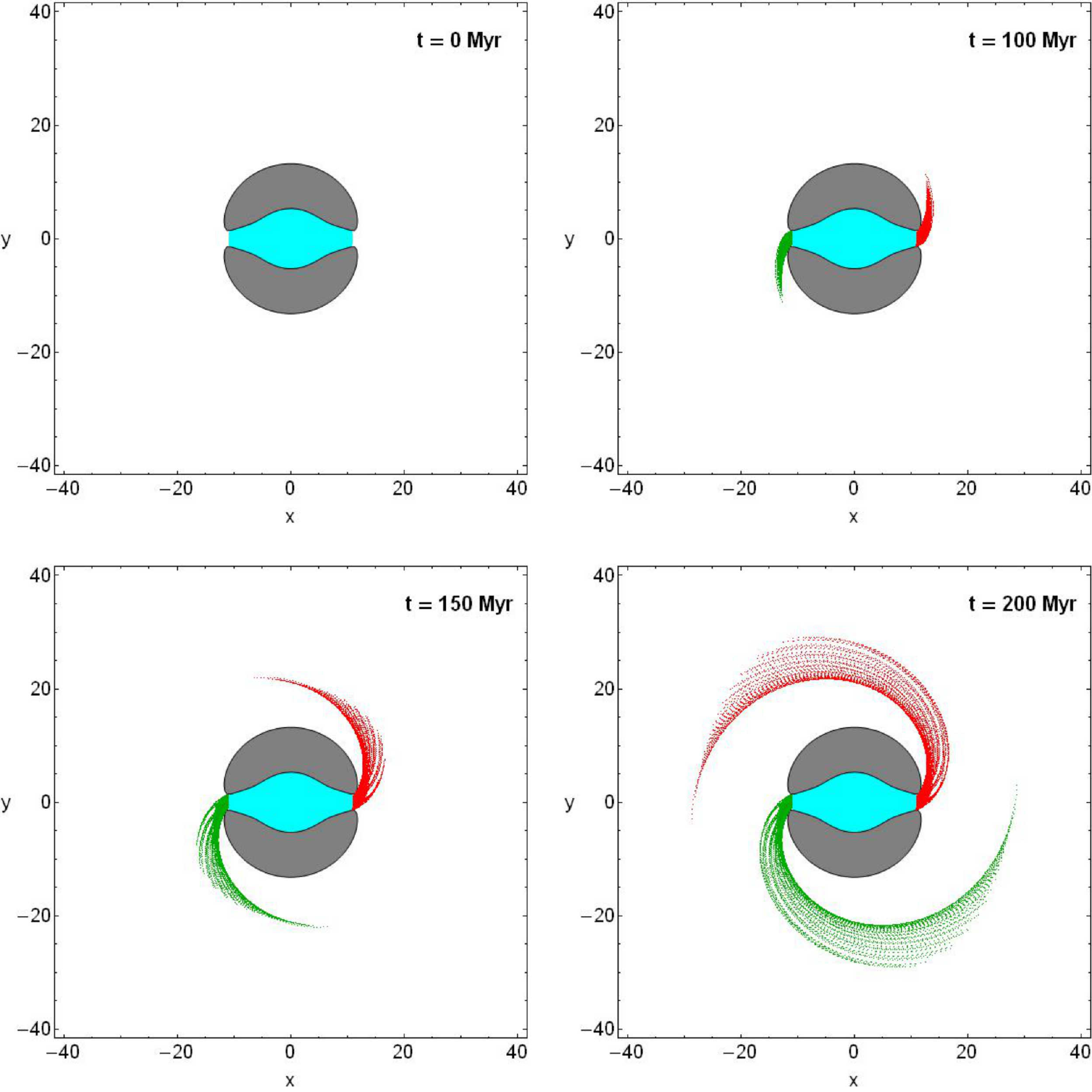}}
\caption{The four parts of the figure show for four different values of the time the distribution of the position of $10^6$ stars in the physical $(x,y)$-plane initiated $(t = 0)$ within the Lagrangian radius, for $E = -1860$ and $\Omega_{\rm b} = -3.5$. We see that as time evolves two symmetrical spiral arms are formed. The green arm contains stars that escaped through $L_1$, while the red arm contains stars that escaped through $L_2$. The bound stars inside the interior galactic region are shown in cyan, while the forbidden regions around $L_4$ and $L_5$ are filled with gray.}
\label{sps}
\end{figure*}

For most of the SALI plots we see a large stable structure at small negative values of $x$ and at large values of $z$. They correspond to tilted loop orbits which encircle the bar along a large loop, where the radial coordinate $R$ is much larger than the width $c_b$ of the bar, and tilted at an angle of approximately $\pi/4$ relative to the axis of the bar. This class of stable orbits seems to be the most persistent class of stable three-dimensional (3D) orbits in our model. Fig. \ref{orb3} shows an example of such an orbit in position space with initial conditions: $x_0 = -3.3$, $y = 0$, $z_0 = 4.45$, $p_{x0} = 0$, $p_{y0} > 0$, $p_{z0} = 0$. The existence of this class of orbits in triaxial rotating models of galaxies is known for a long time, see e.g., \citep{HMS82}.

\section{THE SPIRAL STRUCTURE}
\label{spr}

It is well known that when stars escape from star clusters through the Lagrangian points they form complicated structures known as tidal tails or tidal arms (e.g., \citep{CMM05,dMCM05,JBPE09,KMH08,KKBH10}). In the same vein, in the case of barred spiral galaxies we observe the formation of spiral arms (e.g., \citep{EP14,GKC12,MT97,MQ06,QDB11,RDQ08,SK91}). Usually a star cluster rotates around its parent galaxy in a circular orbit. Thus, the tidal tails follow the curvature defined by the circular orbit and they are nearly straight for orbits moving in large galactocentric distances. The spiral arms of barred galaxies on the other hand, are formed by the non-axisymmetric perturbation of the bar. In particular, in barred spiral galaxies with prominent spiral-like shape the two arms start from the two ends of the bar and then they wind up around the banana-shaped forbidden regions which enclose the Lagrangian points $L_4$ and $L_5$.

It would be very challenging to investigate if our simple gravitational model has the ability to realistically simulate the formation of spiral arms. We shall try to replicate the spiral structure of the SBb galaxy NGC 1300. According to \citet{BT08} (plate 10) the semi-major axis of the bar is about 10 kpc. This means that we have to choose such values of the parameters so that the position of the Lagrangian points $L_1$ and $L_2$ to be at the two ends of the bar. For this purpose we choose $\Omega_{\rm b} = -3.5$, while the values of all the other parameters remain as in SM.

In order to test this we reconfigured our integration numerical routine so as to yield output of all orbit orbits for a three-dimensional grid of size $N_x \times N_y \times N_{\dot{x}} = 100 \times 100 \times 100$, when $z_0 = p_{z0} = 0$. This dense grid of initial conditions is centered at the origin of the physical $(x,y)$ space and we allow for all orbits both signs of $\dot{y}$. Our purpose is to monitor the time-evolution of the path of the orbits and simulate the spiral arm formation. At $t = 0$ all orbits are regularly distributed within the Lagrangian radius inside the interior galactic region. In Fig. \ref{sps} we present four time snapshots of the position of the $10^{6}$ stars in the physical $(x,y)$-plane. We observe that at $t = 1$ time units, which corresponds to 100 Myr the majority of stars are still inside the interior galactic region. However, a small portion of stars has already escaped passing through either $L_1$ or $L_2$. At $t = 1.5$ time units (150 Myr) we have the first indication of the formation of spiral arms, where with green and red color we depict stars that escaped through Lagrangian points $L_1$ and $L_2$, respectively, while stars that remain inside the Lagrangian radius are shown in cyan. As time goes by we see that the two symmetrical spiral arms grow in size and at $t = 2$ time units (200 Myr) the barred galaxy has obtained its complete spiral structure, while the interior galactic region is uniformly filled. In Fig. \ref{sps}(a-d) the density of points along one star orbit is taken to be proportional to the velocity of the star according to \citet{EP14}. Being more specific, a point is plotted (showing the position of a star), if an integer counter variable which is increased by one at every integration step, exceeds the velocity of the star. Following this technique we can simulate, in a way, a real $N$-body simulation of the spiral evolution of the barred galaxy, where the density of stars will be highest where the corresponding velocity is lowest. Therefore, we proved that for $E = -1860$ and $\Omega_{\rm b} = -3.5$ the scenario of evolution of spiral arms in our new barred galaxy model is indeed viable. Additional numerical simulations reveal that this is also true for other (lower or higher) values of energy. In fact, the particular energy level mainly controls how tight the spiral arms wind up around the forbidden regions of motion.

\section{DISCUSSION AND CONCLUSIONS}
\label{disc}

The main topic of the article is a rather simple analytic gravitational model for the potential of a rotating bar in a disk galaxy with an additional spherical dense nucleus. We claim that our new model has some advantages over older models treated in the literature. Our model has intentionally only three components (nucleus, bar, disk) so as to be able to compare the results with the previous model used in \citet{P84}.

Advantages of our model of the bar compared to the model with the Ferrer's triaxial bar: We have a relatively simple functional form with a moderate number of parameters. The parameters $a$ and $c_b$ control the shape of the bar and these parameters have a simple physical meaning: $c_b$ is the width in $y$ and $z$ directions, while $a$ is the length in $x$ direction. If one wants to have different widths in $y$ and $z$ direction then one can replace in Eq. (\ref{monop}) $z$ by $\gamma \ z$ and in analogy in the following equations by inserting this modified function $\upsilon$. In addition we have the parameters $M_{\rm b}$ and $\Omega_{\rm b}$ which are mass and angular velocity of the bar. Fortunately, the function $\Phi_b(x,y,z)$ is an elementary function and therefore it is trivial to include a closed form of the forces into a computer program to construct orbits. In addition we have the same functional form globally. We do not need any cuts at large distance nor any switch to other functional forms. Interestingly, the mass density resulting from our model and as plotted in Fig. \ref{isod} is very similar in shape to the one resulting from the more complicated traditional models and as plotted in Fig. 1 of \citet{P84}. This may be taken as a first hint that our model gives realistic density distributions with moderate effort.

In contrast to simple logarithmic models our new model of the bar has the correct asymptotic behaviour for large distance. As seen in Eq. (\ref{Vb}) also our new model is basically a logarithmic model. However, the argument of the logarithm in Eq. (\ref{Vb}) is constructed such that we obtain the correct asymptotic form $-M_{\rm b}/r$ of the potential for the limit of large distance $r$ from the center. This correct asymptotic behaviour will become relevant when we want to study scattering behaviour, i.e. when we want to study how outside objects become trapped by the galaxy (for most initial conditions only temporarily) and escape again to infinity. Of course, before starting to investigate scattering behaviour one should add to the total potential an additional term describing the halo of the galaxy.

As a further justification of our model serves the behaviour of the most important periodic orbit families of the dynamics. As seen in the Poincar\'{e} sections plotted in section \ref{numres2} one of the stable types of motion in the plane $S_z$ is clockwise (negative orientation) motion around the center with maximal negative angular momentum. Such orbits have their largest distance from the center for $\phi = \pm \pi/2$ but they are rather close to circular. For our typical parameter values the average distance from the center for such orbits is around 6 kpc. Another class of important periodic orbits are the ones of x1 type, as described in detail in subsection \ref{x1o}. They are mainly oscillations along the bar. Such orbits are stable for some parameter regions, in particular for small values of $a$. But interestingly, when they become unstable for more interesting larger values of $a$ then they form a whole infinity of similar descendants forming a braid of such orbits of very moderate global instability. The persistence of this braid of x1 orbits may be taken as a dynamical explanation of the formation and stability of the bar structure.

These two most prominent types of motion would also be of highest importance for the 3-dof dynamics if they would be stable under perturbations out of the invariant plane, i.e. under perturbations in $z$ or $p_z$ direction. We have made some fast preliminary calculations of the behaviour of the periodic orbit for large negative values of $L$ and also for the 1:3 x1 orbit including small perturbations in $p_z$. We have done it for values of $a$ where the respective in plane orbit is still stable. The numerical results indicate that also the orbits with out of plane perturbations remain stable. This numerical behaviour can be taken as evidence that the corresponding types of motion also serve as organizing centers of the dynamics in the full 3-dof dynamics. The most persistent stable orbits for all parameter values of the bar are the inclined 3D loop orbits  where a typical example has been shown in Fig. \ref{orb3}.

The results from section \ref{spr} suggest a stellar dynamical explanation of the formation of spirals which start at the ends of the bar and consist of stars leaking out from the interior part of the effective potential via the Lagrangian points
$L_1$ and $L_2$. Also this behaviour is consistent with the results of \citet{EP14}. Thus we may claim that our new galactic potential has the ability to model the formation and also the evolution of twin spiral structures observed in all barred galaxies.

\section*{ACKNOWLEDGMENTS}

One of the authors (CJ) thanks DGAPA for financial support under grant number IG-101113. The authors would like to express their warmest thanks to the anonymous referee for the careful reading of the manuscript and for all the apt suggestions and comments which allowed us to improve both the quality and the clarity of our paper.


\begin{thebibliography}{}

\bibitem[\protect\citeauthoryear{Athanassoula}{1984}]{A84} Athanassoula, E., 1984, Phys. Rep., 114, 319

\bibitem[\protect\citeauthoryear{Athanassoula}{1992}]{A92} Athanassoula, E,. 1992, MNRAS, 259, 345

\bibitem[\protect\citeauthoryear{Athanassoula}{2003}]{A03} Athanassoula, E., 2003, MNRAS, 341, 1179

\bibitem[\protect\citeauthoryear{Athanassoula et al.}{1983}]{ABMP83} Athanassoula, E., Bienayme, O., Martinet, L., Pfenniger, D., 1983, A\&A, 127, 349

\bibitem[\protect\citeauthoryear{Barazza et al.}{2008}]{BJM08} Barazza, F.D., Jogee, S., Marinova, I., 2008, ApJ, 675, 1194

\bibitem[\protect\citeauthoryear{Benedict et al.}{2002}]{BHJKS02} Benedict, G.F., Howell, D.A., J{\o}rgensen, I., Kenney, J.D.P., Smith, B.J., 2002, AJ, 123, 1411

\bibitem[\protect\citeauthoryear{Berentzen et al.}{2007}]{BSMH07} Berentzen, I., Shlosman, I., Martinez-Valpuesta, I., Heller, C.H., 2007, ApJ, 666, 189

\bibitem[\protect\citeauthoryear{Binney \& Tremaine}{2008}]{BT08} Binney J., Tremaine S., 2008, Galactic Dynamics, Princeton Univ. Press, Princeton, USA

\bibitem[\protect\citeauthoryear{Bountis et al.}{2012}]{BMA12} Bountis, T., Manos, T., Antonopoulos, Ch., 2012, CeMDA, 113, 63

\bibitem[\protect\citeauthoryear{Buta et al.}{2000}]{BTBC00} Buta, R., Treuthardt, P.M., Byrd, G.G., Crocker, D.A., 2000, AJ, 120, 1289

\bibitem[\protect\citeauthoryear{Capuzzo Dolcetta et al.}{2005}]{CMM05} Capuzzo Dolcetta R., Di Matteo P., Miocchi, P., 2005, AJ, 129, 1906

\bibitem[\protect\citeauthoryear{Carpintero \& Aguilar}{1998}]{CA98} Carpintero D.D., Aguilar L.A., 1998, MNRAS, 298, 1

\bibitem[\protect\citeauthoryear{Chapelon et al.}{1999}]{CCD99} Chapelon, S., Contini, T., Davoust, E., 1999, A\&A, 345, 81

\bibitem[\protect\citeauthoryear{Combes et al.}{1990}]{CDFP90} Combes, F., Debbasch, F., Friedli, D., Pfenniger, D., 1990, A\&A, 233, 82	

\bibitem[\protect\citeauthoryear{Comer\'{o}n et al.}{2010}]{CKB10} Comer\'{o}n, S., Knapen, J.H., Beckman, J.E., Laurikainen, E., Salo, H., Mart\'{i}nez-Valpuesta, Buta, R.J., 2010, MNRAS, 402, 2462

\bibitem[\protect\citeauthoryear{Contopoulos \& Grosb{\o}l}{1989}]{CG89} Contopoulos, G., Grosb{\o}ol, P., 1989, A\&AR, 1, 261

\bibitem[\protect\citeauthoryear{Dalcanton et al.}{2004}]{DYB04} Dalcanton, J.J., Yoachim, P., Bernstein, R.A., 2004, ApJ, 608, 189

\bibitem[\protect\citeauthoryear{Debattista et al.}{2006}]{DMC06} Debattista, V.P., Mayer, L., Carollo, C.M., Moore, B., Wadsley, J., Quinn, T., 2006, ApJ, 645, 209

\bibitem[\protect\citeauthoryear{Di Matteo et al.}{2005}]{dMCM05} Di Matteo P., Capuzzo Dolcetta R., Miocchi P., 2005, CeMDA, 91, 59

\bibitem[\protect\citeauthoryear{Elmegreen \& Elmegreen}{1985}]{EE85} Elmegreen, B.G., Elmegreen, D.M., 1985, ApJ, 288, 438

\bibitem[\protect\citeauthoryear{Englmaier \& Gerhard}{1997}]{EG97} Englmaier, P., Gerhard, O., 1997, MNRAS, 287, 57

\bibitem[\protect\citeauthoryear{Ernst \& Peters}{2014}]{EP14} Ernst A., Peters T., 2014, MNRAS, 443, 2579

\bibitem[\protect\citeauthoryear{Eskridge et al.}{2000}]{Ee00} Eskridge, P.B., Frogel, J.A., Pogge, R.W., Quillen, A.C., Davies, R.L., DePoy, D.L., Houdashelt, M.L., et al., 2000, AJ, 119, 356

\bibitem[\protect\citeauthoryear{Ferrers}{1877}]{F77} Ferrers, N.M. 1877, Q. J. Pure Appl. Math., 14:1

\bibitem[\protect\citeauthoryear{Foyle et al.}{2008}]{FCT08} Foyle, K., Courteau, S., Thacker, R.J., 2008, MNRAS, 386, 1821

\bibitem[\protect\citeauthoryear{Gadotti}{2011}]{G11} Gadotti, D.A., 2011, MNRAS, 415, 3308

\bibitem[\protect\citeauthoryear{Grand et al.}{2012}]{GKC12} Grand R.J.J., Kawata D., Cropper M., 2012, MNRAS, 421, 1529

\bibitem[\protect\citeauthoryear{Heisler et al.}{1982}]{HMS82} Heisler, J., Merritt, D., Schwarzchild, M., 1982, ApJ, 258, 490

\bibitem[\protect\citeauthoryear{Hoyle et al.}{2011}]{HMN11} Hoyle, B., Masters, K.L., Nichol, R.C., Edmondson, E.M., Smith, A.M., Lintott, C., Scranton, R., Bamford, S., Schawinski, K., Thomas, D., 2011, MNRAS, 415, 3627

\bibitem[\protect\citeauthoryear{Hsieh et al.}{2011}]{HMLHOW11} Hsieh, P.Y., Matsushita, S., Liu, G. Ho, P.T.P., Oi, N., Wu, Y.L., 2011, ApJ, 736, 129

\bibitem[\protect\citeauthoryear{Just et al.}{2009}]{JBPE09} Just A., Berczik P., Petrov M., Ernst A., 2009, MNRAS, 392, 969

\bibitem[\protect\citeauthoryear{Kaufmann \& Contopoulos}{1996}]{KC96} Kaufmann, D.E., Contopoulos, G., 1996, A\&A, 309, 381

\bibitem[\protect\citeauthoryear{Kaufmann \& Patsis}{2005}]{KP05} Kaufmann, D., Patsis, P., 2005, ApJ, 624, 693

\bibitem[\protect\citeauthoryear{Kim et al.}{2012a}]{KSSYT12} Kim, W.T., Seo, W.Y., Stone, J.M., Yoon, D., Teuben, P.J., 2012a, ApJ, 747, 60

\bibitem[\protect\citeauthoryear{Kim et al.}{2012b}]{KSK12} Kim, W.T., Seo, W.Y., Kim, Y., 2012b, ApJ, 758, 14

\bibitem[\protect\citeauthoryear{Knapen et al.}{1995}]{KBHSd95} Knapen, J.H., Beckman, J.E., Heller, C.H., Shlosman, I., de Jong, R.S., 1995, ApJ, 454, 623

\bibitem[\protect\citeauthoryear{Kormendy}{1979}]{K79} Kormendy J., 1979, ApJ, 227, 714

\bibitem[\protect\citeauthoryear{Kormendy \& Kennicutt}{2004}]{KK04} Kormendy, J., Kennicutt, R.C., Jr., ARA\&A, 42, 603

\bibitem[\protect\citeauthoryear{K\"{u}pper et al.}{2008}]{KMH08} K\"{u}pper A.H.W., Macleod A., Heggie D.C., 2008, MNRAS, 387, 1248

\bibitem[\protect\citeauthoryear{K\"{u}pper et al.}{2010}]{KKBH10} K\"{u}pper A.H.W., Kroupa P., Baumgardt H., Heggie D.C., 2010, MNRAS, 401, 105

\bibitem[\protect\citeauthoryear{Laine et al.}{2002}]{LSKP02} Laine, S., Shlosman, I., Knapen, J.H., Peletier, R.F., 2002, ApJ, 567, 97

\bibitem[\protect\citeauthoryear{Laurikainen \& Salo}{2002}]{LS02} Laurikainen, E., Salo, H., 2002, MNRAS, 337, 1118

\bibitem[\protect\citeauthoryear{Laurikainen et al.}{2002}]{LSR02} Laurikainen, E., Salo, H., Rautiainen, P., 2002, MNRAS, 331, 880

\bibitem[\protect\citeauthoryear{Laurikainen et al.}{2007}]{LSBK07} Laurikainen, E., Salo, H., Buta, R., Knapen, J.H., 2007, MNRAS, 381, 401

\bibitem[\protect\citeauthoryear{Maciejewski et al.}{2002}]{MTSS02} Maciejewski, W., Teuben, P.J., Sparke, L.S., Stone, J.M., 2002, MNRAS, 329, 502

\bibitem[\protect\citeauthoryear{Manos \& Athanassoula}{2011}]{MA11} Manos, T., Athansssoula, E., 2011, MNRAS, 415, 629

\bibitem[\protect\citeauthoryear{Manos et al.}{2013}]{MBS13} Manos, T., Bountis, T., Skokos, Ch., 2013, J. Phys. A: Math. Theor. 46, 254017

\bibitem[\protect\citeauthoryear{Masset \& Tagger}{1997}]{MT97} Masset F., Tagger M., 1997, A\&A, 322, 442

\bibitem[\protect\citeauthoryear{Masters et al.}{2011}]{MNH11} Masters, K.L., Nichol, R.C., Hoyle, B., Lintott, C., Bamford, S.P., Edmondson, E.M., Fortson, L., Keel, W.C., Schawinski, K., Smith, A.M., Thomas, D., 2011, MNRAS, 411, 2026

\bibitem[\protect\citeauthoryear{Mazzuca et al.}{2008}]{MKVR08} Mazzuca, L.M., Knapen, J.H., Veilleux, S., Regan, M.W., 2008, ApJ, 174, 337

\bibitem[\protect\citeauthoryear{Minchev \& Quillen}{2006}]{MQ06} Minchev I., Quillen A.C., 2006, MNRAS, 368, 623

\bibitem[\protect\citeauthoryear{Miyamoto \& Nagai}{1975}]{MN75} Miyamoto W., Nagai R., 1975, PASJ 27, 533

\bibitem[\protect\citeauthoryear{Oll\'{e} \& Pfenniger}{1998}]{OP98} Oll\'{e}, M., Pfenniger, D., 1998, A\&A, 334, 829

\bibitem[\protect\citeauthoryear{P\'{e}rez \& S\'{a}nchez-Bl\'{a}zquez}{2011}]{PSB11} P\'{e}rez, I., S\'{a}nchez-Bl\'{a}zquez, P., 2011, A\&A, 529, A64

\bibitem[\protect\citeauthoryear{Pfenniger}{1984}]{P84} Pfenniger D., 1984, A\&A 134, 373

\bibitem[\protect\citeauthoryear{Pfenniger}{1996}]{P96} Pfenniger, D., 1996, in Buta R., Crocker D. A., Elmegreen B. G. eds, ASP Conf. Ser. Vol. 91, Barred Galaxies. Astron. Soc. Pac., San Francisco, p. 273

\bibitem[\protect\citeauthoryear{Pichardo et al.}{2004}]{PMM04} Pichardo, B., Martos, M., Moreno, E., 2004, ApJ, 609, 144

\bibitem[\protect\citeauthoryear{Piner et al.}{1995}]{PST95} Piner, B.G., Stone, J.M., Teuben, P.J., 1995, ApJ, 449, 508

\bibitem[\protect\citeauthoryear{Press et al.}{1992}]{PTVF92} Press, H.P., Teukolsky, S.A, Vetterling, W.T., Flannery, B.P., 1992, Numerical Recipes in FORTRAN 77, 2nd Ed., Cambridge University Press, Cambridge, USA

\bibitem[\protect\citeauthoryear{Quillen et al.}{2011}]{QDB11} Quillen A.C., Dougherty J., Bagley M.B., Minchev I., Comparetta J., 2011, MNRAS, 417, 762

\bibitem[\protect\citeauthoryear{Regan \& Elmegreen}{1997}]{RE97} Regan, M.W., Elmegreen, D.M., 1997, AJ, 114, 965

\bibitem[\protect\citeauthoryear{Regan \& Teuben}{2003}]{RT03} Regan, M.W., Teuben, P.J., 2003, ApJ, 582, 723

\bibitem[\protect\citeauthoryear{Ro\v{s}kar et al.}{2008}]{RDQ08} Ro\v{s}kar R., Debattista V.P., Quinn T.R., Stinson G.S., Wadsley J., 2008, ApJ, 684, L79

\bibitem[\protect\citeauthoryear{S\'{a}nchez-Bl\'{a}zquez et al.}{2011}]{SBO11} S\'{a}nchez-Bl\'{a}zquez, P., Ocvirk, P., Gibson, B.K., P\'{e}rez, I., Peletier, R.F., 2011, MNRAS, 415, 709

\bibitem[\protect\citeauthoryear{Sandstrom et al.}{2010}]{Se10} Sandstrom, K.; Krause, O., Linz, H., Schinnerer, E., Dumas, G., Meidt, S., Rix, H.W., Sauvage, M., Walter, F., Kennicutt, R.C., et al., 2010, A\&A, 518, 59

\bibitem[\protect\citeauthoryear{Sellwood \& Kahn}{1991}]{SK91} Sellwood J.A., Kahn F.D., 1991, MNRAS, 250, 278

\bibitem[\protect\citeauthoryear{Sellwood \& Wilkinson}{1993}]{SW93} Sellwood, J., Wilkinson, A., 1993, Rep. Prog. Phys., 56, 173

\bibitem[\protect\citeauthoryear{Sheth et al.}{2003}]{SRSS03} Sheth, K., Regan, M.W., Scoville, N.Z., Strubbe, L.E., 2003, ApJ, 592, L13

\bibitem[\protect\citeauthoryear{Skokos}{2001}]{S01} Skokos, C., 2001, Journal of Physics A, 34, 10029

\bibitem[\protect\citeauthoryear{Skokos et al.}{2002a}]{SPA02a} Skokos, Ch., Patsis, P.A., Athanassoula, E., 2002a MNRAS, 333 847

\bibitem[\protect\citeauthoryear{Skokos et al.}{2002b}]{SPA02b} Skokos, Ch., Patsis, P.A., Athanassoula, E., 2002n, MNRAS, 333, 861

\bibitem[\protect\citeauthoryear{Thakur et al.}{2009}]{TAJ09} Thakur, P., Ann, H.B., Jiang, I., 2009, ApJ, 693, 586

\bibitem[\protect\citeauthoryear{Wada \& Koda}{2001}]{WK01} Wada, K., Koda, J., 2001, PASJ, 53, 1163

\bibitem[\protect\citeauthoryear{Weinzirl et al.}{2009}]{WJK09} Weinzirl, T., Jogee, S., Khochfar, S., Burkert, A., Kormendy, J., 2009, ApJ, 696, 411

\bibitem[\protect\citeauthoryear{Zotos \& Caranicolas}{2013}]{ZC13} Zotos, E.E., Caranicolas, N.D, 2013, Nonlinear Dynamics, 74, 1203

\bibitem[\protect\citeauthoryear{Zotos}{2014}]{Z14} Zotos, E.E., 2014, Nonlinear Dynamics, 76, 1301

\end{thebibliography}
\end{document}